\documentclass{jfm}
\usepackage{graphicx}
\usepackage{epstopdf, epsfig}

\usepackage{amsmath}
\usepackage[colorlinks, citecolor=blue]{hyperref}
\usepackage{color}
\usepackage{soul}
\usepackage{setspace}

\usepackage{lipsum}
\usepackage{siunitx}

\shorttitle{Taylor-Culick retractions}
\shortauthor{V. Sanjay, U. Sen, P. Kant, and D. Lohse}

\title{Taylor-Culick retractions and the influence of the surroundings}

\author{Vatsal Sanjay\aff{1}
  \thanks{Both authors contributed equally to this work.}
  \corresp{\email{vatsalsanjay@gmail.com}},
  Uddalok Sen\aff{1}
  \footnotemark[1]
  \corresp{\email{u.sen@utwente.nl}},
  Pallav Kant\aff{1},
 \and Detlef Lohse{\aff{1}$^{,}$\aff{2}}
 \corresp{\email{d.lohse@utwente.nl}}}

\affiliation{\aff{1}Physics of Fluids Group, Max Planck Center for Complex Fluid Dynamics, Department of Science and Technology, MESA+ Institute for Nanotechnology, and J. M. Burgers Centre for Fluid Dynamics, University of Twente, P. O. Box 217, 7500 AE Enschede, The Netherlands
\aff{2}Max Planck Institute for Dynamics and Self-Organization, Am Fassberg 17, 37077 G\"{o}ttingen, Germany}

\newcommand{\Wef}{\mathit{We}_\mathit{f}}
\newcommand{\Cas}{\mathit{Ca}_\mathit{s}}
\newcommand{\Ohs}{\mathit{Oh}_\mathit{s}}
\newcommand{\Ohf}{\mathit{Oh}_\mathit{f}}
\newcommand{\Oha}{\mathit{Oh}_\mathit{a}}
\newcommand{\gammaaf}{\gamma_\mathit{af}}
\newcommand{\gammasf}{\gamma_\mathit{sf}}
\newcommand{\gammasa}{\gamma_\mathit{sa}}

\begin{document}
\maketitle

\begin{abstract}
When a freely suspended liquid film ruptures, it retracts spontaneously under the action of surface tension. If the film is surrounded by air, the retraction velocity is known to approach the constant Taylor-Culick velocity. However, when surrounded by an external viscous medium, the dissipation within that medium dictates the magnitude of the retraction velocity. In the present work, we study the retraction of a liquid (water) film in a viscous oil ambient (\emph{two-phase} Taylor-Culick retractions), and that sandwiched between air and a viscous oil (\emph{three-phase} Taylor-Culick retractions). In the latter case, the experimentally-measured retraction velocity is observed to have a weaker dependence on the viscosity of the oil phase as compared to the configuration where the water film is surrounded completely by oil. Numerical simulations indicate that this weaker dependence arises from the localization of viscous dissipation near the three-phase contact line. The speed of retraction only depends on the viscosity of the surrounding medium and not on that of the film. From the experiments and the numerical simulations, we reveal unprecedented regimes for the scaling of the Weber number $\Wef$ of the film (based on its retraction velocity) or the capillary number $\Cas$ of the surroundings vs. the Ohnesorge number $\Ohs$ of the surroundings in the regime of large viscosity of the surroundings ($\Ohs \gg 1$), namely $\Wef \sim \Ohs^{-2}$ and $\Cas \sim \Ohs^{0}$ for the two-phase Taylor-Culick configuration, and $\Wef \sim \Ohs^{-1}$ and $\Cas \sim \Ohs^{1/2}$ for the three-phase Taylor-Culick configuration. 
\end{abstract}

\begin{keywords}

\end{keywords}

\section{Introduction}
Liquid films, sheets, and shells have piqued the interest of fluid dynamicists for nearly two centuries \citep{savart1833OppositeJets, savart1833WaterBells, savart1833HydraulicJump, taylor1959dynamics, taylor-1959-procrsoclonda, clanet2007waterbells, villermaux2020fragmentation}. A freely suspended liquid film warrants further attention among these configurations as it is inherently metastable owing to its high surface area. Indeed, if a large enough \citep{taylor1973making, villermaux2020fragmentation} hole nucleates on the film, the sheet will spontaneously retract to reduce its surface area. Such interfacial destabilization leading to film rupture and bursting can also result in waterborne disease transmission \citep{bourouiba2021fluid}. The bursting of liquid films at an oil-air interface is important for various industrial applications in the chemical and petrochemical engineering sectors as well. One area of particular interest is underwater oil spills in oceans, such as the Deepwater Horizon spill in 2010 in the Gulf of Mexico~\citep{summerhayes2011deep}. For these spills, droplets (or slugs) of oil may rise to the free surface of water via buoyancy, and then rupture the free surface of water directly above it. The water film will retract upon rupture, and the oil will spread on the water surface, thus perpetuating an environmental hazard. 

\begin{figure}
\centering
\includegraphics[width=\textwidth]{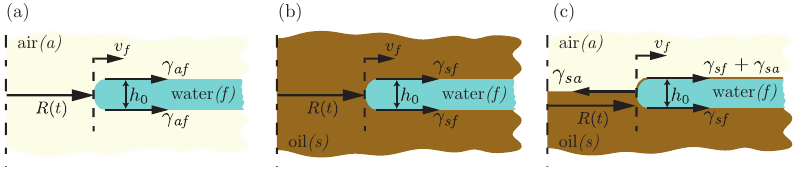}
\caption{Schematics depicting the configurations studied in the present work: (a) retraction of a water film ($f$) of thickness $h_0$ in an air ($a$) environment (\emph{classical} configuration),  (b) retraction of water film ($f$) in an oil ($s$) environment (\emph{two-phase} configuration), (c) retraction of a water film sandwiched between air and oil (\emph{three-phase} configuration). The dot-dashed line represents the axis of rotational symmetry, and $R(t)$ is the radius of the growing hole centered at this axis. In all the schematics, the water film is retracting from left to right with velocity $v_f$, as indicated by the arrow, and $\gamma_{ij}$ denotes the surface tension coefficient between fluids $i$ and $j$.}
\label{fig:configs}
\end{figure}

Perhaps the most widely studied example of sheet destabilization and retraction is during the bursting of liquid (e.g., soap) films in air -- an area of active research since the pioneering works \citep{dupre1867theorie, dupre1869theorie, rayleigh-1891-nature, taylor-1959-procrsoclonda, culick-1960-japplphys, mcentee-1969-jphyschem} in the late nineteenth and mid-twentieth century to the more recent investigations \citep{bremond-2005-jfm, muller-2007-prl, lhuissier-2012-jfm, munro-2015-jfm, deka-2020-prf}. In these studies, the outer medium is assumed passive (inviscid and zero-inertia). The origin of the nucleation of the initial hole in the film can be manifold~\citep{lohse-2020-pnas}. After film rupture, the internal viscous stresses in the film do not contribute to the momentum balance, but dictate the distribution of momentum within the film \citep{savva-2009-jfm}, as long as the Ohnesorge number of the film (ratio of its visco-capillary to inertio-capillary time scales, see \S~\ref{sec::Num method}) is less than its aspect ratio \citep[see][]{deka-2020-prf}. Nonetheless, half of the surface energy released goes into internal viscous dissipation (see appendix~\ref{App::ClassicalTC} and \citet{culick-1960-japplphys,de1996introductory, sunderhauf-2002-pof, villermaux2020fragmentation}). 

A representative schematic of the situation mentioned above is shown in figure~\ref{fig:configs}a (henceforth referred to as the \emph{classical} Taylor-Culick configuration), where the water film ($f$) of thickness $h_0$ is retracting in air ($a$) under the action of surface tension. The retraction velocity, $v_f$, in such a scenario is constant (after a period of initial transience) and approaches the Taylor-Culick velocity given by

\begin{align}
v_{\text{TC}} = \sqrt{\frac{2\gammaaf}{\rho_f h_0}} ,
\label{eq:v_TC}
\end{align}

\noindent where $2\gammaaf$ is the net surface tension driving the retraction ($\gammaaf$ being the interfacial tension coefficient between the film and air) and $\rho_f$ is the density of the liquid film (figure~\ref{fig:configs}a). Furthermore, it was observed that during the retraction, the liquid collects in a thicker rim at the retracting edge of the film, particularly for low viscosity liquids \citep[not depicted in figure~\ref{fig:configs}a]{rayleigh-1891-nature, ranz-1959-japplphys, pandit-1990-jfm}. The seminal work of \citet{keller-1983-pof} further explores the retraction of these films of non-uniform thickness. 

The effect of viscosity of the film ($\eta_f$) during its retraction process has also been studied \citep{debregeas-1995-prl, debregeas-1998-science}. \citet{brenner-1999-pof} showed that although viscosity does not have any effect on the constant retraction velocity, it can have a significant effect on the shape of the retracting edge of a planar film. They report that if the radial extent of the film is greater than its Stokes length ($= \eta_f/(\rho_f v_{\text{TC}})$), a growing rim is formed, whereas the rim is absent for the converse situation. \citet{savva-2009-jfm} extended the work by \citet{brenner-1999-pof} for highly viscous films, and also developed a lubrication model for the retraction dynamics of a circular hole. They concluded that although viscosity does not determine the magnitude of the constant retraction velocity, it does dictate the time required (post rupture) to attain that constant velocity, which increases with increasing viscosity. Recently, \citet{pierson2020revisiting, deka-2020-prf} revisited the viscous retraction dynamics by exploring self-similar solutions for slender filaments and sheets of finite length. 

The rheological properties of the film also influence the retraction dynamics \citep{kdv-1999-pre, tammaro-2018-langmuir, villone-2019-jnnfm, kamat-2020-jfm}. For instance, \cite{sen_lohse_2021} showed that viscoelastic filaments can retract at velocities higher than the Newtonian Taylor-Culick limit owing to elastic tension. Moreover, the retraction dynamics of liquids have also been studied in the context of dewetting for a wide range of scenarios \citep{redon-1991-prl, brochardwyart-1993-langmuir, shull-1994-langmuir, andrieu-1996-jadhesion, lambooy-1996-prl, haidara-1998-langmuir, buguin-1999-prl, peron-2012-langmuir, peschka-2018-scirep, kim-2020-pnas}. Lastly, there has also been a recent surge in the study of liquid retraction in other configurations, such as liquid strips \citep{lv-2015-jfm}, smectic films \citep{trittel-2013-pof}, foam films \citep{petit-2015-jfm}, and emulsion films \citep{vernay-2015-prl}. 

In all the aforementioned studies, the surrounding medium is assumed to play no role in the rupture dynamics. A question naturally arises: what happens when the outer medium also interacts with the retracting film? In particular, how do the viscosity and the inertia of the outer medium influence the rupture dynamics \citep{mysels-1973-jphyschem, joanny-1987-physicaa, reyssat-2006-epl, jian_deng_thoraval_2020}? A representative schematic for such a scenario is shown in figure~\ref{fig:configs}b, where a water film is retracting in a viscous oil ambient. This geometry will henceforth be referred to as the \emph{two-phase} configuration where the net surface tension force responsible for retraction is $2\gammasf$ ($\gammasf$ being the interfacial tension coefficient between the film and the surrounding medium, see figure~\ref{fig:configs}b). In such a situation, viscous dissipation is not limited only to the retracting film, but is also present in the ambient. If the ambient happens to be significantly more viscous than the film, then the dissipation in the ambient dominates. In such a situation, the retraction velocity is still a constant. However, unlike the classical case, the velocity depends on the viscosity $\eta_{s}$ of the ambient medium \citep{martin-1994-epl, reyssat-2006-epl}. Common realizations of this configuration include relaxation of filaments and droplets in a viscous medium \citep{stone1989relaxation}, or that of air-films during drop impact \citep{jian2020split, jian_deng_thoraval_2020}. Additionally, in this context, \cite{anthony2020initial} showed that the so-called \lq\lq inertially limited viscous regime\rq\rq\, in the early-times of drop coalescence \citep{paulsen-2012-pnas, paulsen2013approach} stems from a Taylor-Culick type retraction of the air film between the deformable drops.

In the present work, we study the influence of the surrounding medium on the retraction velocity of a ruptured liquid film using both force balance and energy conservation arguments. To accomplish this goal, along with the two canonical configurations shown in figures~\ref{fig:configs}a and~\ref{fig:configs}b, we also study the retraction dynamics of a liquid film sandwiched between air and a viscous oil bath. A representative schematic is shown in figure~\ref{fig:configs}c. This geometry will henceforth be referred to as the \emph{three-phase} configuration. This paper elucidates this case experimentally by inflating an oil droplet at the water-air interface and letting the water film rupture. Such a configuration can also be found in the early stages of water film retraction when an air bubble approaches a water-oil interface if the oil-layer is thick enough \citep{feng2014nanoemulsions, feng_muradoglu_kim_ault_stone_2016}. Furthermore, we also use direct numerical simulations (DNS) to demystify the retraction dynamics by using a precursor film-based three-fluid volume of fluid (VoF) method. We show that the film in this three-phase configuration still retracts with a constant velocity, and similar to the two-phase case, the retraction velocity depends on the viscosity $\eta_{s}$ of the oil bath. However, this dependence is weaker in the three-phase configuration. Furthermore, we reveal an unprecedented scaling relationship for the retraction velocity of the film, which arises from the localization of the viscous dissipation near the three-phase contact line. 

The paper is organized as follows: \S~\ref{sec:3-phaseTC Exp methods} describes the problem statement for the three-phase Taylor-Culick retractions along with the experimental method employed to probe this configuration. The results from these experiments are discussed in \S~\ref{sec:3-phaseTC Exp results}. \S~\ref{sec::Num method} presents the numerical framework, and \S~\ref{sec:Num results} describes the simulation results for both the two-phase and three-phase configurations. \S~\ref{sec:forces} demonstrates the balance of forces in Taylor-Culick retractions, followed by the corresponding scaling relationships in \S~\ref{sec:scaling}. Further, \S~\ref{sec:energetics} analyzes the overall energy balance, highlighting the differences in the viscous dissipation mechanisms between the two-phase and three-phase configurations. The work culminates with conclusions in \S~\ref{sec:conclusion}. Throughout the manuscript, we refer to Appendix~\ref{App::ClassicalTC} for discussions on the classical Taylor-Culick retractions, and use the experimental datapoints from \citet{reyssat-2006-epl} for the two-phase configuration.

\section{Film bursting at an air-liquid interface: experimental method}\label{sec:3-phaseTC Exp methods}
\begin{figure}
	\centering
	\includegraphics[width=\textwidth]{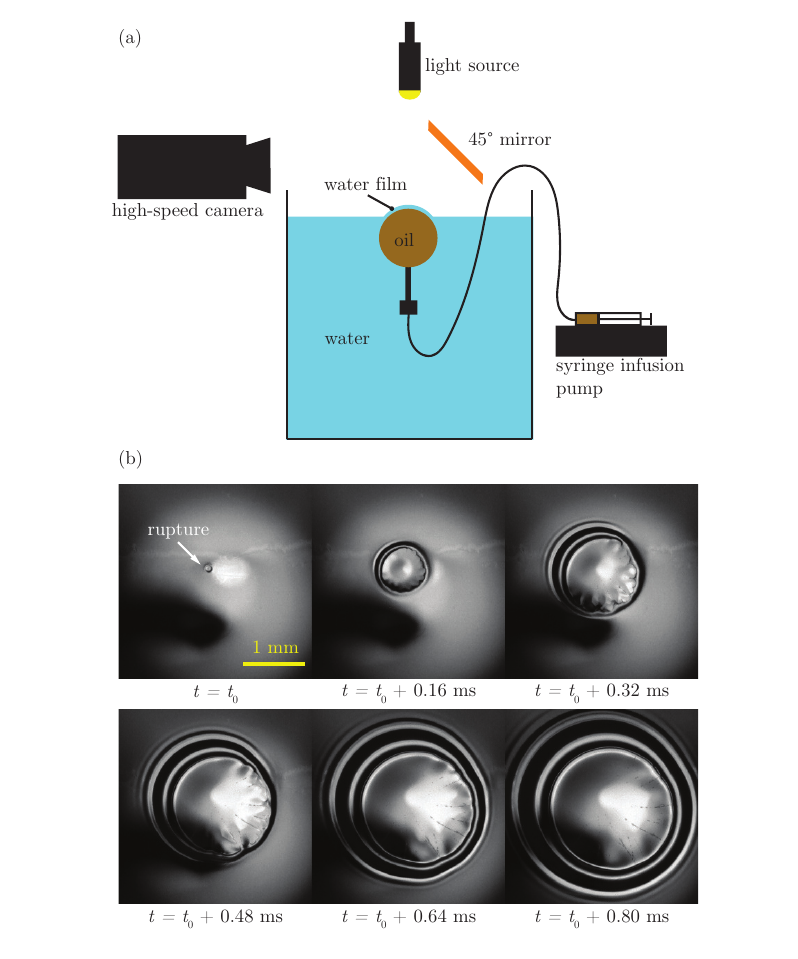}
	\caption{(a) Schematic of the experimental setup; (b) typical time-lapsed experimental snapshots of the film rupture and the subsequent retraction process ($\nu_{s}$ = 10 cSt). The time instant $t$ = $t_{0}$ denotes the first frame where rupture (indicated by the white arrow) is discernible.}
	\label{fig:setup}
\end{figure}

We study the three-phase configuration experimentally by inflating an oil drop (`$s$' for `surroundings' in figure~\ref{fig:configs}c) at a water-air free interface and capturing the retraction of the water film ($f$ in figure~\ref{fig:configs}c). The schematic of the experimental setup is shown in figure~\ref{fig:setup}a. A plastic box of dimensions $25\,\si{\milli\meter}\,\times\,25\,\si{\milli\meter}\,\times15\,\si{\milli\meter}$ (length $\times$ width $\times$ height, Bodemschat) filled with purified water (Milli-Q) was used as the liquid bath for most of the experiments. To study the effect of the viscosity of the retracting film, the water in the bath was replaced by glycerol (Sigma-Aldrich)-water mixtures (concentrations in the range 50\% -- 70\% by wt.) for some experiments. A dispensing needle (inner diameter = $0.41\,\si{\milli\meter}$, HSW Fine-Ject) was submerged within the bath such that its dispensing end was at a depth of $2.4\,\si{\milli\meter}$ from the free surface (depth kept constant during all experiments). A silicone oil (Wacker) droplet was created at the tip of the needle by connecting it to an oil-filled plastic syringe ($5\,\si{\milli\liter}$, Braun Injekt) via a flexible plastic PEEK tubing (Upchurch Scientific). The oil flow rate was maintained at $0.05\,\si{\milli\liter}/\si{\minute}$ with the help of a syringe infusion pump (Harvard Apparatus). In the present experiments, silicone oils of different viscosities were used, and their densities ($\rho_{s}$), kinematic ($\nu_{s}$), and dynamic viscosities ($\eta_{s}$) are listed in Table~\ref{tab:prop}. It is to be noted that for only the AK 0.65 oil ($\eta_{s} = 4.94 \times 10^{-4}\,\si{\pascal}.\si{\second}$), the drop is less viscous than the water film ($\eta_{f} = 8.9 \times 10^{-4}\,\si{\pascal}.\si{\second}$), while for all the other oils, the film is less viscous. The oil-water interfacial tension ($\gamma_{sf}$) was considered to be $0.040\,\si{\newton}/\si{\meter}$ \citep{peters-2013-colloidssurfa}. 

The drop volume was increased by a slow infusion using the syringe pump. The needle depth below the free surface was chosen such that the drop remained anchored to the needle during inflation. As a result, the water film right above the oil droplet progressively thins with increasing volume of the drop. Below a certain thickness, the film ruptures due to van der Waals forces \citep{vaynblat-2001-pof}, and subsequently retracts into the bath. This situation is analogous to the rupture and retraction of a water film sandwiched between air and a viscous oil droplet. This is also a configuration that is flipped vertically as compared to the early time scenario studied by \citet{feng_muradoglu_kim_ault_stone_2016}. Another key difference is that we ensure negligible vertical velocity at the point of rupture of the water film, making this scenario ideal for studying three-phase Taylor Culick retractions.

High-speed imaging of this rupture and retraction phenomena was performed at 50000~fps~(frames-per-second) for the lower viscosity oils and at 10000~fps for the higher viscosity ones, with a $2.5\,\si{\micro\second}$ exposure time, by a high-speed camera (Fastcam Nova S12, Photron) connected to a macro lens (DG Macro $105\,\si{\milli\meter}$, Sigma) with $64\,\si{\milli\meter}$ of lens extender (Kenko). The camera was pointed at a plane mirror (Thorlabs) inclined at $45^{\circ}$ to the horizontal to capture the top view of the retraction phenomenon~(figure~\ref{fig:setup}a), while the experiments were illuminated from the top by a LED light source (KL 2500 LED, Schott). A typical bursting event is shown in figure~\ref{fig:setup}b, where time $t$~=~$t_{0}$ indicates the instant when rupture is optically discernible. With increasing time, the size of the hole formed due to rupture increases as the film retracts. The phenomenological observations shown in figure~\ref{fig:setup}b will be discussed in detail in \S~\ref{sec:3-phaseTC Exp results}. The captured images were then further analyzed using the open-source software FIJI \citep{fiji} and an in-house OpenCV-based Python script to obtain quantitative information presented in the following sections.

\begin{table}
  \begin{center}
\def~{\hphantom{0}}
  \begin{tabular}{lccc}
      Silicone oil & $\rho_{s}$ (kg/m$^{3}$) & $\nu_{s}$ (cSt) & $\eta_{s}$ (Pa.s)\\[3pt]
       AK 0.65 & 760 & 0.65 & 4.94 $\times$ 10$^{-4}$\\
       AK 5 & 920 & 5 & 4.60 $\times$ 10$^{-3}$\\
       AK 10 & 930 & 10 & 9.30 $\times$ 10$^{-3}$\\
       AK 20 & 950 & 20 & 1.90 $\times$ 10$^{-2}$\\
       AK 35 & 960 & 35 & 3.36 $\times$ 10$^{-2}$\\
       AK 50 & 960 & 50 & 4.80 $\times$ 10$^{-2}$\\
       AK 100 & 960 & 100 & 9.60 $\times$ 10$^{-2}$\\
       AK 200 & 970 & 200 & 1.94 $\times$ 10$^{-1}$\\
       AK 350 & 970 & 350 & 3.40 $\times$ 10$^{-1}$\\
       AK 1000 & 970 & 1000 & 9.70 $\times$ 10$^{-1}$\\
  \end{tabular}
  \caption{Salient properties of the silicone oils used in the present work.}
  \label{tab:prop}
  \end{center}
\end{table}

\section{Film bursting at an air-liquid interface: experimental results}\label{sec:3-phaseTC Exp results}
\begin{figure}
\centering
\includegraphics[width=\textwidth]{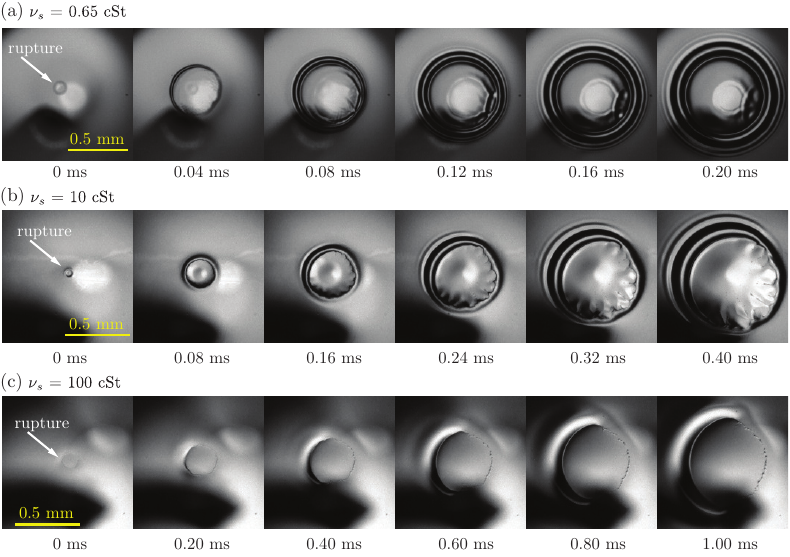}
\caption{Time-lapsed snapshots of the post-rupture retraction of water films for different oil viscosities: (a) $\nu_{s}$ = 0.65 cSt, (b) $\nu_{s}$ = 10 cSt, and (c) $\nu_{s}$ = 100 cSt. The rupture location is denoted by the white arrows, while the time stamps indicate the time since rupture is first observed (i.e., $t - t_{0}$). Also see supplementary movie SM1.}
\label{fig:exp snapshots}
\end{figure}

The rupture and retraction of a water film on the surface of an oil drop of $\eta_{s}~=~4.94~\times~10^{-4}~\si{\pascal}.\si{\second}$ is shown in figure~\ref{fig:exp snapshots}a (and supplementary movie SM1). The timestamps indicate ($t - t_{0}$), where $t_{0}$ is the time instant when rupture is optically discernible, and $t$ is the current time. As mentioned earlier, in this particular case, the water film is more viscous than the oil. It is to be noted that in the present experiments, we could not precisely control the location of rupture as it was sensitive to experimental noise \citep[see \S~4.2 of][]{villermaux2020fragmentation}. Hence, the rupture in the present experiments did not always occur at the apex of the thinning film. Such behavior was also observed in other similar experiments of film rupture \citep{oldenziel-2012-pof, demaleprade-2016-prl}. The rupture location may also be determined by a `prehole' formation, also observed by \citet{vernay-2015-prl} for the bursting of emulsion-based liquid sheets. In their work, \citet{vernay-2015-prl} show that the presence of emulsion oil droplets at the air-water interface results in lowering of the local interfacial tension, leading to Marangoni flows away from that location. This flow leads to a local thinning of the film, which ultimately ruptures at that location. They also report that the prehole formation always precedes rupture in their experiments. In the present experiments, the water surface is never pristine and always contains small impurities (which are practically unavoidable). It is possible that these impurities might have reduced the local surface tension, resulting in a similar Marangoni flow leading to a prehole. For the discussion on the origin of the hole nucleation, we also refer to \citet{lohse-2020-pnas}. In any case, upon rupture, a circular hole is formed in the film, which grows radially in time. Therefore, the oil bounded by the periphery of the hole gets into contact with air and not with the water film. It is also noticeable that the edge of the retracting film forms a thick rim -- an observation also made for the retraction of liquid films in air \citep{pandit-1990-jfm, brenner-1999-pof, sunderhauf-2002-pof}. The presence of the rim can be qualitatively surmised from the experimental snapshots (figures \ref{fig:setup}b, \ref{fig:exp snapshots}a, and \ref{fig:exp snapshots}b), where the change in the curvature of the film downstream of the rim introduces a difference in the color intensity. As the hole increases in size (or as the film retracts further), this rim also becomes thicker. Finally, since silicone oil prefers to spread on water \citep{li-2020-pnas}, the retraction process ceases when the oil droplet has completely spread on water, thus creating a macroscopic film whose thickness is controlled by volume conservation and thermodynamics \citep{book-degennes}. 

When the viscosity of the oil phase is increased to $\eta_{s} = 9.30 \times 10^{-3}\,\si{\pascal}.\si{\second}$, the hole opening (or film retraction) dynamics (as seen in figure~\ref{fig:exp snapshots}b and supplementary movie SM1) are qualitatively similar to that for the lower viscosity described before (figure~\ref{fig:exp snapshots}a). Here also, the film forms a thick rim at its retracting edge. Nonetheless, an increase in the oil's viscosity decreases the retraction speed, as indicated by the timestamps (corresponding to $t - t_{0}$). This behavior is expected since the physical situation is analogous to a retracting water film shearing the free surface of viscous oil: increasing $\eta_{s}$ increases the resistance to shearing, which in turn makes the retraction process slower. 

Furthermore, the experimental snapshots show that the oil-air-water contact line exhibits corrugations during the retraction process, and fine streams of droplets are released from these corrugations. Similar observations were also made for film retraction in the two-phase configuration \citep{reyssat-2006-epl, oldenziel-2012-pof} and during the rupture of the intermediate film when a drop coalesces with a pool of the same liquid in the presence of an external medium \citep{aryafar-2008-pre, kavehpour-2015-arfm}. The nature of the corrugations is reminiscent of the sharp tips observed during selective withdrawal \citep{cohen-2002-prl, courrechdupont-2006-prl, courrechdupont-2020-pnas} or tip streaming \citep{montanero-2020-rpp}. In a frame of reference co-moving with the rim, the film sees highly viscous oil being aspirated away from it, resulting in the formation of the sharp tips. Indeed, such a mechanism was also hinted at by \citet{reyssat-2006-epl} for the instabilities observed in their experiments for film retraction in the two-phase configuration. \citet{tseng-2015-jfm} showed that such instabilities are formed due to the local convergence of streamlines in the neighborhood of a zero-vorticity point or line on the interface. However, a detailed and quantitative investigation of the formation and subsequent breakup of these liquid tips is beyond the scope of the present work. 

\begin{figure}
\centering
\includegraphics[width=\textwidth]{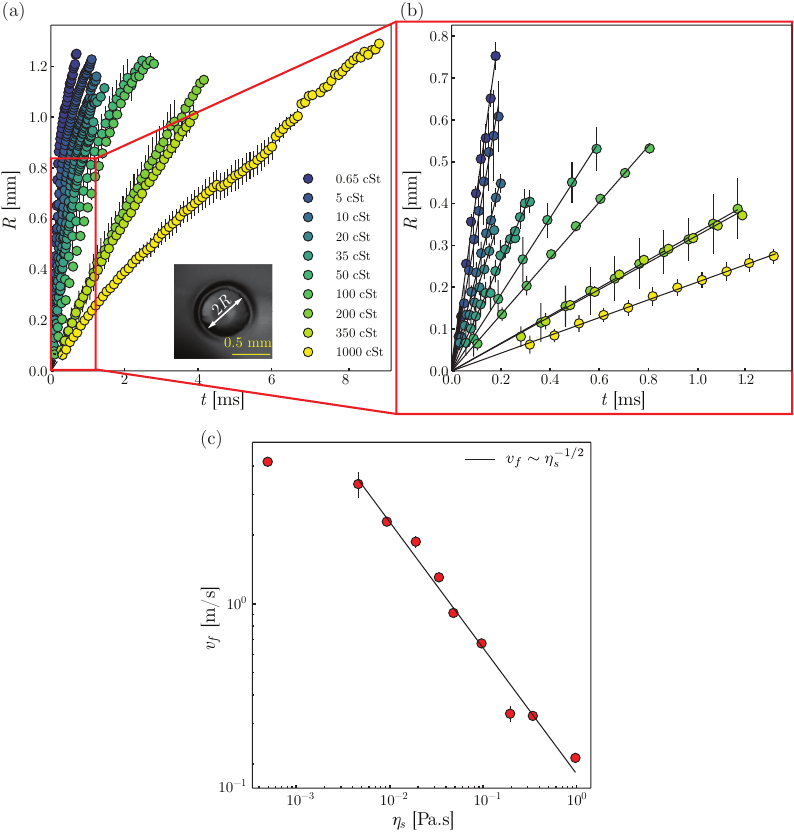}
\caption{(a) Temporal evolution of the retraction radius ($R$) for oils of different kinematic viscosity ($\nu_{s}$); a typical measurement is shown in the snapshot in the inset. (b) At early times (red rectangle in figure~\ref{fig:dynamics}a), $R$ varies linearly with time; the discrete datapoints are experimental measurements and the lines are linear fits. (c) Variation of dewetting velocity ($v_f$) with the dynamic viscosity of the oil phase ($\eta_{s}$); the discrete datapoints are experimental measurements and the line represents $v_f \sim \eta_{s}^{-1/2}$.}
\label{fig:dynamics}
\end{figure}

For an even higher viscosity of the oil phase ($\eta_{s} = 9.60 \times 10^{-2}\,\si{\pascal}.\si{\second}$, see figure~\ref{fig:exp snapshots}c and supplementary movie SM1), the retraction of the ruptured water film is further slowed down (as evident from the timestamps in figure~\ref{fig:exp snapshots}c). Furthermore, the retracting edge also does not possess a thick rim. This observation is similar to the case of \citet{brenner-1999-pof} for the retraction of viscous films in air, where films of higher viscosity do not form a rim. Moreover, although the expanding holes for the lower $\eta_{s}$ cases (as shown in figures~\ref{fig:exp snapshots}a and \ref{fig:exp snapshots}b) are almost circular, the one for the high viscosity case shown in figure~\ref{fig:exp snapshots}c is highly asymmetric. This asymmetry can be attributed to the location of the rupture not being at the film's apex. Since the rupture is happening at an off-apex location, the film thickness at the location of rupture is not spatially uniform due to the curvature of the oil droplet. Hence, the retraction velocity is faster on the part of the film which has a lower thickness. Presumably, this effect is more pronounced when the overall film retraction dynamics are slower, as is the case for the experiments shown in figure~\ref{fig:exp snapshots}c. To confirm this hypothesis, one requires high-resolution measurements of the spatial variation of the film thickness, which is challenging in the present experiments (further discussed in \S~\ref{sec:scaling}). The corrugations at the oil-air-water contact line are also observed in this case. However, since the retraction velocity itself is considerably smaller than for the case shown in figure~\ref{fig:exp snapshots}b (see figure~\ref{fig:dynamics}b for specific values), the tips are not as sharp, and no droplet streams are observed. 

To quantify the retraction dynamics, we measure the hole opening radius from each snapshot captured using the high-speed camera. For each experimental snapshot, the area of the hole $A(t)$ is measured, and subsequently an equivalent hole opening radius $R(t)$ is calculated as $A(t) = \pi (R(t))^{2}$. A typical measurement from the optical images is depicted in the inset of figure~\ref{fig:dynamics}a. The temporal variation of the measured hole radius, $R$, is shown in figure~\ref{fig:dynamics}a. The time instant corresponding to the first frame in which rupture is optically discernible is denoted by $t_{0}$. Each datapoint in figure~\ref{fig:dynamics}a denotes the mean of measurements from five independent experiments, and the error bars correspond to $\pm$ one standard deviation. In the present work, we focus on the early moments following rupture, as indicated by the red rectangle in figure~\ref{fig:dynamics}a. Zooming into this early time regime, as shown in figure~\ref{fig:dynamics}b, it is observed that $R(t)$ varies linearly with time (as evident from the lines denoting linear fits in figure~\ref{fig:dynamics}b). This variation indicates that the retraction velocity $v_f$ (= $dR/dt$), given by the slopes of the linear fits, is constant for each viscosity. This is reminiscent of the constant rupture velocity also observed for the classical (figure~\ref{fig:configs}a) and two-phase (figure~\ref{fig:configs}b) Taylor-Culick configurations. Furthermore, it is also observed that with increasing $\eta_{s}$ (or $\nu_{s}$), the slope of the linear fits (hence $v_f$) decreases, as expected from the qualitative observations reported in figure~\ref{fig:exp snapshots}. 

The variation of $v_f$ with $\eta_{s}$ is shown in figure~\ref{fig:dynamics}c. The typical retraction velocities are~$\mathcal{O}\left(1\,\si{\meter}/\si{\second}\right)$. A decreasing $v_f$ with increasing $\eta_{s}$ is observed. Furthermore, for the cases where the oil is more viscous than water ($\eta_{f} = 8.9 \times 10^{-4}\,\si{\pascal}.\si{\second}$), the retraction velocity varies as

\begin{align}
v_{f} \sim \frac{1}{\eta_{s}^{1/2}}, 
\label{eq:exp scaling}
\end{align}

\noindent as evident from the line in figure~\ref{fig:dynamics}c. This is a weaker dependence as compared to the expected $1/\eta_{s}$ variation observed for retraction in the two-phase configuration \citep{martin-1994-epl, eri-2010-pre}. We will attempt to explain the scalings for the two-phase and three-phase configurations in \S~\ref{sec:scaling}. Furthermore, the reason for not fitting the datapoint for the case where the oil is less viscous than the water film in figure~\ref{fig:dynamics}c will also be addressed therein. 

\section{Numerical framework}\label{sec::Num method}
\subsection{Governing equations}
\begin{figure}
	\centering
	\includegraphics[width=\textwidth]{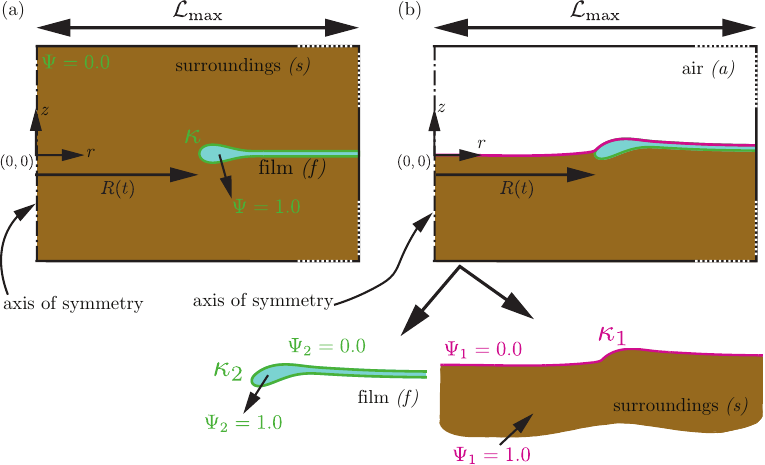}	
	\caption{Computational domain for (a) two-phase and (b) three-phase Taylor-Culick retractions. For the classical case, (a) is used by replacing the surroundings ($s$) with air ($a$). The size of the domain is much larger than the hole radius $(\mathcal{L}_\text{max} \gg R(t))$. Furthermore, $\mathcal{L}_\text{max}/h_0 \gg \text{max}\left(\Ohf, \Ohs\right)$.}
	\label{fig:NumericalSetup}
\end{figure}

In this section, we discuss the governing equations that describe the retraction of a ruptured liquid film in the three configurations we study in this paper, namely, the classical, two-phase, and three-phase Taylor-Culick retractions. We perform axisymmetric direct numerical simulations using the free-software volume of fluid (VoF) program, \emph{Basilisk C} \citep{popinet-basilisk, basiliskVatsal}, which uses the one-fluid approximation \citep{tryggvason2011direct} to solve the continuity and the Navier-Stokes equations:

\begin{align}
	\label{Eqn::continuity}
	\boldsymbol{\nabla\cdot v} &= 0,\\ 
	\label{Eqn::NS}
	\rho\left(\frac{\partial\boldsymbol{v}}{\partial t} + \boldsymbol{\nabla\cdot}\left(\boldsymbol{vv}\right)\right) &= -\boldsymbol{\nabla} p + \boldsymbol{\nabla\cdot}\left(2\eta\boldsymbol{\mathcal{D}}\right) + \boldsymbol{f}_\gamma
\end{align}

\noindent where, $\boldsymbol{v}$ and $p$ are the velocity vector and pressure fields, respectively, $\eta$ is the viscosity of the fluid, and $t$ denotes time. Furthermore, $\boldsymbol{\mathcal{D}}$ is the symmetric part of the velocity gradient tensor $\left(\boldsymbol{\mathcal{D}} = \left(\boldsymbol{\nabla v} + \left(\boldsymbol{\nabla v}\right)^{\text{T}}\right)/2\right)$, and $\boldsymbol{f}_\gamma$ is the singular surface tension force needed in the one-fluid approximation to comply with the dynamic boundary condition at the interfaces \citep{brackbill1992continuum}. 

\subsection{Non-dimensionalization of the governing equations}
We non-dimensionalize the governing equations by using the inertio-capillary velocity scale $v_\gamma$, the thickness of the film $h_0$, and the capillary pressure $p_\gamma$. These scales also define the characteristic inertio-capillary time as $\tau_\gamma$:

\begin{align}
	\label{Eqn::Scales}
	\tau_\gamma = \frac{h_0}{v_\gamma} = \sqrt{\frac{\rho_fh_0^3}{2\gammasf}},  \quad v_\gamma = \sqrt{\frac{2\gammasf}{\rho_f h_{0}}}, \quad p_\gamma = \frac{2\gammasf}{h_0}.
\end{align}

\noindent Here, $\gammasf$ is the surface tension coefficient between the film ($f$) and the surrounding ($s$) medium, $\rho_f$ the film density, and $h_0$ its thickness. The dimensionless form of the Navier-Stokes equation (\ref{Eqn::NS}) is

\begin{align}
	\label{Eqn::NS2}
	\tilde{\rho}\left(\frac{\partial\boldsymbol{\tilde{v}}}{\partial \tilde{t}} + \boldsymbol{\tilde{\nabla}\cdot}\left(\boldsymbol{\tilde{v}}\boldsymbol{\tilde{v}}\right)\right) = -\boldsymbol{\tilde{\nabla}} p + \boldsymbol{\tilde{\nabla}\cdot}\left(2Oh\boldsymbol{\tilde{\mathcal{D}}}\right) + \boldsymbol{\tilde{f}}_\gamma,
\end{align}

\noindent where the expressions for the Ohnesorge number ($Oh$, ratio of visco-capillary to inertio-capillary time scales), the dimensionless density ($\tilde{\rho}$), and the singular surface tension force ($\boldsymbol{\tilde{f}}$) depend on the specific configurations that we discuss below.  

\subsubsection{Two-phase Taylor-Culick configuration}\label{sec:2-phaseTC Num setup}

In this configuration, a liquid film ($f$) retracts in a viscous surrounding ($s$) medium~(figure~\ref{fig:NumericalSetup}a). We use the volume of fluid (VoF) tracer $\Psi$ to differentiate between the film~($\Psi~=~1$) and the surroundings ($\Psi = 0$), which follows the VoF scalar advection equation, 

\begin{align}
	\label{Eqn::Vof1}
	\left(\frac{\partial}{\partial\tilde{t}} + \boldsymbol{\tilde{v}\cdot\tilde{\nabla}}\right)\Psi = 0.
\end{align}

\noindent Furthermore, the singular surface tension force is given by \citep{brackbill1992continuum}:

\begin{align}\label{Eqn::SurfaceTension1}
	\boldsymbol{\tilde{f}}_\gamma \approx \left(\tilde{\kappa}/2\right)\boldsymbol{\tilde{\nabla}}\Psi,
\end{align}

\noindent where the curvature $\kappa$ is calculated using the height-function approach \citep{popinet-2009-jcp}. We follow the same sign convention as \citet[][see page 33]{tryggvason2011direct}: the curvature is positive if the interface folds towards it normal $\boldsymbol{\hat{n}}$, i.e., $\kappa = - \boldsymbol{\nabla \cdot n}$. Note that the surface tension scheme in Basilisk C is explicit in time. So, we restrict the maximum time step as the characteristic inertio-capillary time based on the wavelength of the smallest capillary wave. Additionally, the density of the film is the same as that of the surroundings, giving $\tilde{\rho} = 1$. Lastly, the Ohnesorge number ($Oh$) is given by

\begin{align}
	\label{Eqn::Oh2}
	Oh = \Psi \Ohf + \left(1-\Psi\right)\Ohs,
\end{align}

\noindent where

\begin{align}
	\label{Eqn::Ohs}
	\Ohf = \frac{\eta_f}{\sqrt{\rho_f\left(2\gammasf\right)h_0}} \quad \text{and} \quad \Ohs = \frac{\eta_s}{\sqrt{\rho_f\left(2\gammasf\right)h_0}}
\end{align}

\noindent are the Ohnesorge numbers based on the film and surroundings viscosities, respectively. For this configuration, we keep $\Ohf$ constant at $0.05$ \citep[based on the experiments of][]{reyssat-2006-epl}, and vary the control parameter $\Ohs$ in \S~\ref{sec:Num results}. 

Note that the computational domain in figure~\ref{fig:NumericalSetup}a along with (\ref{Eqn::Vof1}) -- (\ref{Eqn::Ohs}) can be used to simulate classical Taylor-Culick retractions as well by replacing the surroundings~($s$) with air ($a$). We discuss the details of the classical configuration in appendix~\ref{App::ClassicalTC}. 

\subsubsection{Three-phase Taylor-Culick configuration}\label{sec:3-phaseTC Num methods}
In this configuration, we model the bursting of a water film at an oil drop-air interface by simulating the retraction of a fluid film ($f$) on an initially flat oil bath ($s$), while ignoring the effects of the oil drop's curvature (as the retraction length in the early time regime of figure~\ref{fig:dynamics}b is much smaller than the oil drop radius, see figure~\ref{fig:NumericalSetup}b). We extend the traditional volume of fluid~(VoF) method described in \S~\ref{sec:2-phaseTC Num setup} to tackle three fluids by using two VoF tracers: $\Psi_1$, which is tagged as $1$ for the liquids (water film,  $f$, and oil surroundings,  $s$) and $0$ for air ($a$), and $\Psi_2$ which is $1$ for the water film ($f$) and $0$ everywhere else (figure~\ref{fig:NumericalSetup}b). Note that this implementation requires an implicit declaration of the surrounding phase ($s$), given by $\Psi_2\left(1-\Psi_1\right)$ \citep{sanjay2019droplet, vatsal-basilisk-3p, basiliskVatsal, mou2021singular}. Additionally, both $\Psi_1$ and $\Psi_2$ follow the VoF tracer advection equation,

\begin{align}
	\label{Eqn::Vof2}
	\left(\frac{\partial}{\partial\tilde{t}} + \boldsymbol{\tilde{v}\cdot\tilde{\nabla}}\right)\{\Psi_1, \Psi_2\} = 0,
\end{align}

\noindent and the dimensionless density ratio is (with $\rho_f = \rho_s$)

\begin{align}
	\label{Eqn::density3}
	\tilde{\rho} = \Psi_1 + \left(1-\Psi_1\right)\left(\rho_a/\rho_f\right).
\end{align}

\noindent The Ohnesorge number ($Oh$) is now given by

\begin{align}
	\label{Eqn::Oh3}
	Oh = \Psi_1\Psi_2\Ohf + \left(1-\Psi_2\right)\Psi_1\Ohs + \left(1-\Psi_1\right)\Oha,
\end{align}

\noindent where $\Ohf$ and $\Ohs$ follow (\ref{Eqn::Ohs}), and $\Oha = \eta_a/\sqrt{\rho_f\left(2\gammasf\right)h_0}$ is the Ohnesorge number based on the viscosity of air. Both $\Ohf$ and $\Oha$ are fixed at $10^{-1}$ and $10^{-3}$, respectively, for all the three-phase simulation data presented in this paper (see \S~\ref{sec:scaling}), and we vary the control parameter $\Ohs$ in \S~\ref{sec:Num results}. Lastly, the surface tension body force takes the form

\begin{align}\label{Eqn::SurfaceTension2}
	\boldsymbol{\tilde{f}}_\gamma \approx \left(\gammasa/\gammasf\right)\left(\tilde{\kappa}_1/2\right)\boldsymbol{\tilde{\nabla}}\Psi_1 + \left(\tilde{\kappa}_2/2\right)\boldsymbol{\tilde{\nabla}}\Psi_2,
\end{align}

\noindent with $\gammasa$ and $\gammasf$ being the surface tension coefficients for the surroundings-air and surroundings-film interfaces, respectively. 

Physically, such a configuration (figure~\ref{fig:NumericalSetup}b) and (\ref{Eqn::Vof2}) -- (\ref{Eqn::Oh3}) ideally imply the presence of a zero thickness precursor film of the surrounding liquid ($s$, represented by $(1-\Psi_2)\Psi_1 = 1$, (\ref{Eqn::Oh3})) over the liquid film ($f$, $\Psi_1\Psi_2 = 1$, (\ref{Eqn::Oh3})). Note that this numerical assumption is applicable only when it is thermodynamically favorable for one of the fluids (here $s$) to spread over all the other fluids, i.e., it has a positive spreading coefficient \citep{book-degennes, berthier2012physics}, $S \equiv \gammaaf - \gammasf - \gammasa > 0$, and the Neumann triangle collapses at the three-phase contact line. In reality, this precursor film will have a finite thickness controlled by microscopic forces \citep[like van der Waals forces,][]{vaynblat-2001-pof}, and is much smaller than the length scales that we can resolve numerically in the continuum framework. Indeed, for our numerical simulations, this precursor film has an effective thickness of $\Delta /2$, where $\Delta$ is the size of the finest grid employed in this work. We further assume that, on the time scale of film retraction, the effective spreading coefficient of the surrounding liquid ($s$) is $0$ \citep{bonn2009wetting}. Consequently, the effective surface tension coefficient between the film and air is $\gammaaf = \gammasf+\gammasa$. This precursor film \citep{thoraval-2013-pre} is analogous to the mathematical model for spreading of a perfectly wetting liquid on a solid substrate \citep{book-degennes, bonn2009wetting}, which regularizes the contact line singularity owing to the numerical slip \citep[with an effective slip length of $\Delta$/2,][]{afkhami2018transition} due to the discretization of the interface.

\subsection{Note on non-dimensionalization in the viscous regime}\label{sec::ViscousScaling}
For highly viscous surroundings ($\Ohs > 1$), it is convenient to scale the velocities with the visco-capillary velocity scale $v_\eta$, owing to the dominant interplay between viscous and capillary stresses \citep{stone1989relaxation}. Further, we can use the visco-capillary time $\tau_\eta$, film thickness $h_0$, and capillary pressure $p_\gamma$ to normalize the time, length, and pressure dimensions, respectively:

\begin{align}
	\label{Eqn::Scales2}
	\tau_\eta = \frac{h_0}{v_\eta} = \frac{\eta_s h_0}{2\gammasf}, \quad v_\eta = \frac{2\gammasf}{\eta_s}, \quad p_\gamma = \frac{2\gammasf}{h_0},
\end{align}

\noindent where $\gammasf$ is the surface tension coefficient between the film ($f$) and the surroundings~($s$), $h_0$ the film thickness, and $\eta_s$ the viscosity of the surrounding medium. These visco-capillary scales modify the momentum equation as

\begin{align}
	\label{Eqn::NS3}
	\frac{\tilde{\rho}}{Oh_s^2}\left(\frac{\partial\boldsymbol{\tilde{v}}}{\partial \tilde{t}} + \boldsymbol{\tilde{\nabla}\cdot}\left(\boldsymbol{\tilde{v}}\boldsymbol{\tilde{v}}\right)\right) = -\boldsymbol{\tilde{\nabla}} p + \boldsymbol{\tilde{\nabla}\cdot}\left(2\tilde{\eta}\boldsymbol{\tilde{\mathcal{D}}}\right) + \boldsymbol{\tilde{f}}_\gamma.
\end{align}

\noindent Here, $\Ohs$ is the surroundings Ohnesorge number (\ref{Eqn::Ohs}), $\tilde{\rho}$ follows $\tilde{\rho} = 1$ and (\ref{Eqn::density3}) for the two-phase and the three-phase configurations, respectively, and $\boldsymbol{\tilde{f}}_\gamma$ equals the corresponding expressions for the two configurations. Additionally, the dimensionless viscosities are given by

\begin{align}
	\label{Eqn::Oh4}
	\tilde{\eta} = 
	\begin{cases}
		\Psi \left(\eta_f/\eta_s\right) + \left(1-\Psi\right)&\text{two-phase case},\\
		\\
		\Psi_1\Psi_2\left(\eta_f/\eta_s\right) + \left(1-\Psi_2\right)\Psi_1 + \left(1-\Psi_1\right)\left(\eta_a/\eta_s\right)&\text{three-phase case}.
	\end{cases}
\end{align}

\subsection{Domain size and boundary conditions}
Figure~\ref{fig:NumericalSetup} depicts the computational domains. The left boundary represents the axis of symmetry with origin marked at $(0, 0)$. We set no-penetration and free-slip boundary conditions to all other domain boundaries along with zero gradient conditions for pressure. These boundaries are far away from the expanding hole and do not affect its growth. Furthermore, the size of the domain is chosen such that $\mathcal{L}_{\text{max}} \gg \text{max}\left(\Ohf, \Ohs\right)$, with a minimum $\mathcal{L}_{\text{max}}$ of $200$ for $\Ohs \ll 1$. We have varied this domain size to ensure that the simulations are independent of its value. Note that, if this condition is not met, the assumption of infinite film, which is essential for the theoretical scaling relations developed in this work, will fail \citep{deka-2020-prf}. 

We employ Adaptive Mesh Refinement (AMR) to correctly resolve the different interfaces as well as regions of high velocity gradients (and hence, high viscous dissipation, see appendix~\ref{App::EnergyBalance}). To ensure that the velocity field is captured accurately, these refinement criteria \citep[see][]{basiliskVatsal} effectively maintain a minimum of $40$ cells across the thickness of the film (i.e., $h_0/\Delta \ge 40$). As the apparent three-phase contact line and the viscous boundary layer are critical in the present work, the refinement criteria maintain a minimum of $40$ cells in the wedge region near the apparent three-phase contact line. Furthermore, the viscous boundary layer is almost $10$ times larger than the film thickness (see \S~\ref{sec:energetics viscous}, figure~\ref{fig:dissipation}). Consequently, a minimum of $400$ cells in the viscous boundary layer in the surrounding medium is needed to properly resolve the velocity gradients. We have conducted extensive grid independence studies so that the final results (energy transfers and the retraction velocity) are independent of the number of grid cells. 

\section{Taylor-Culick retractions: numerics}\label{sec:Num results}

\begin{figure}
	\centering
	\includegraphics[width=0.925\textwidth]{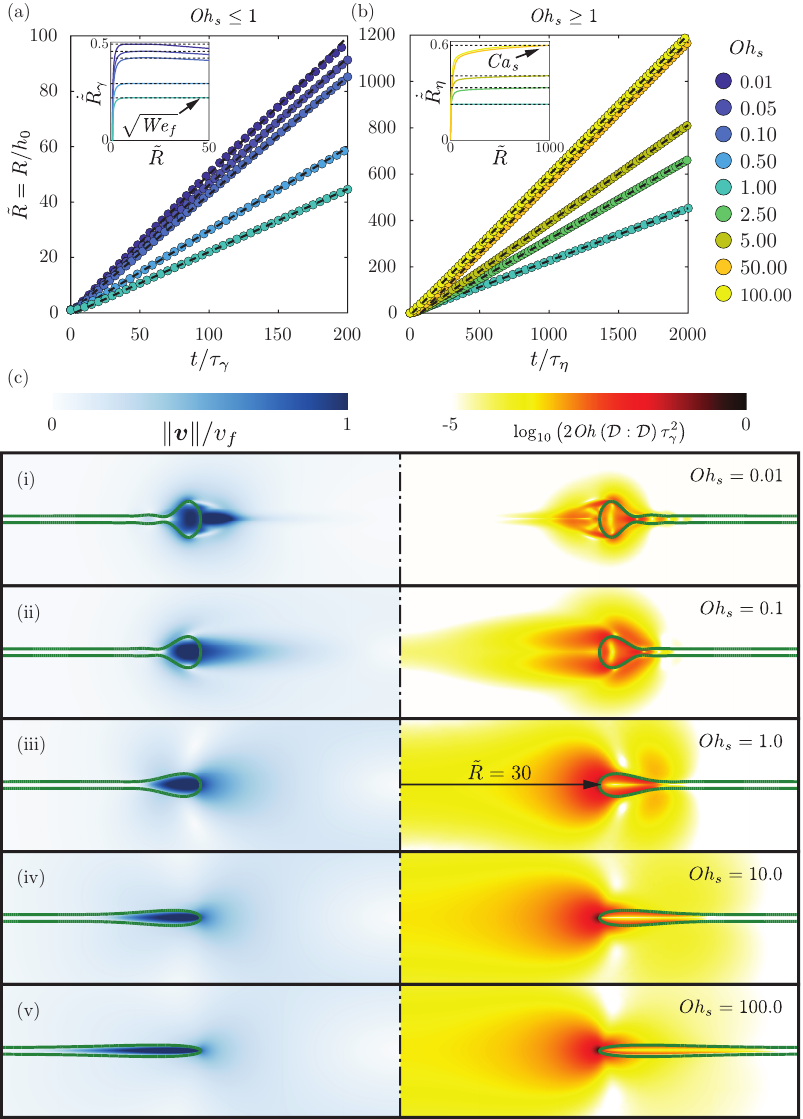}	
	\caption{Two-phase Taylor-Culick retractions: temporal evolution of the dimensionless hole radius ($\tilde{R}$($t$)) for (a) $\Ohs \le 1$ and (b) $\Ohs \ge 1$. Time is normalized using the inertio-capillary time scale, $\tau_\gamma = \sqrt{\rho_f h_0^3/\gammasf}$ in panel (a) and the visco-capillary time scale, $\tau_\eta = \eta_s h_0/\gammasf$ in panel (b). Insets of these panels show the variation of the dimensionless growth rate of the hole radius at different $\Ohs$, and mark the definitions of $\Wef$ and $\Cas$. Lastly, panel (c) illustrates the morphology of the flow at different $\Ohs$ at $\tilde{R} = 30$. In each snapshot, the left hand side contour shows the velocity magnitude normalized with the (terminal) film velocity $v_f$ and the right hand side shows the dimensionless rate of viscous dissipation per unit volume normalized using the inertio-capillary scales, represented on a $\log_{\text{10}}$ scale to differentiate the regions of maximum dissipation. Here, the film Ohnesorge number is $\Ohf = 0.05$. Also see supplementary movie SM2.}
	\label{fig:two-phaseTemporal}
\end{figure}

\begin{figure}
	\centering
	\includegraphics[width=0.925\textwidth]{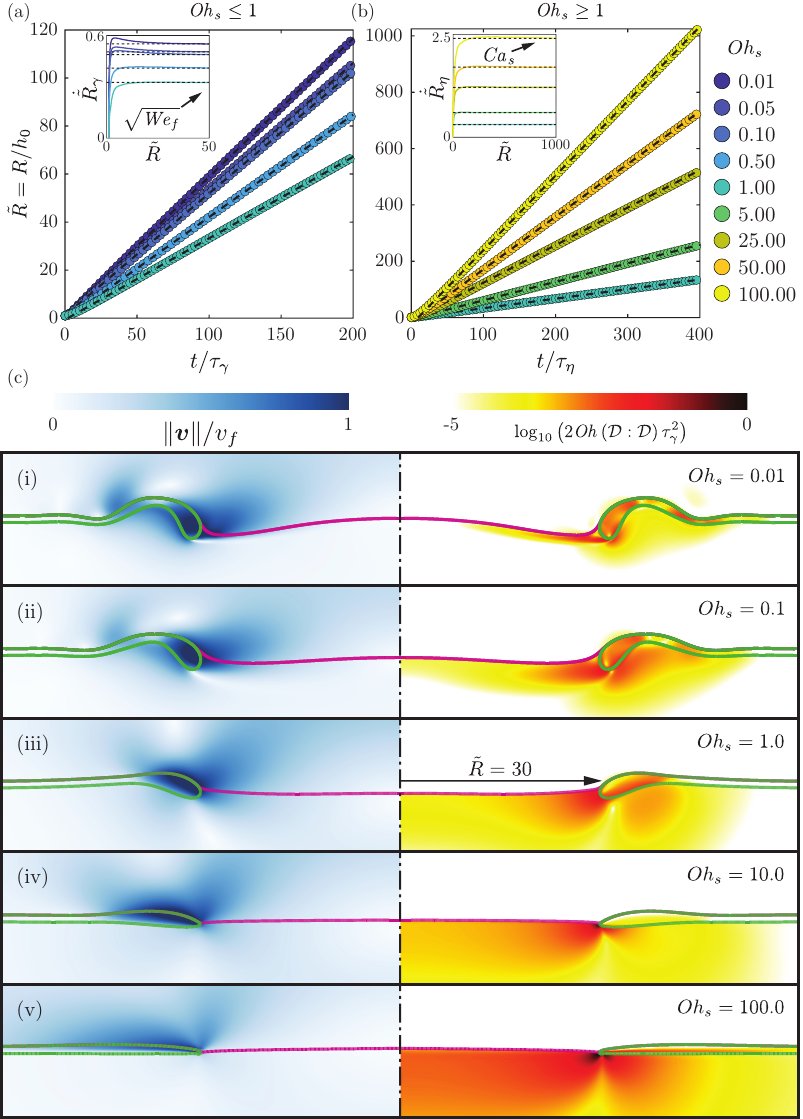}	
	\caption{Three-phase Taylor-Culick retractions: temporal evolution of the dimensionless hole radius ($\tilde{R}(t)$) for (a) $\Ohs \le 1$ and (b) $\Ohs \ge 1$. Time is normalized using the inertio-capillary time scale, $\tau_\gamma = \sqrt{\rho_f h_0^3/\gammasf}$ in panel (a) and the visco-capillary time scale, $\tau_\eta = \eta_s h_0/\gammasf$ in panel (b). Insets of these panels show the variation of the dimensionless growth rate of the hole radius at different $\Ohs$, and mark the definitions of $\Wef$ and $\Cas$. Lastly, panel (c) illustrates the morphology of the flow at different $\Ohs$ at $\tilde{R} = 30$. In each snapshot, the left hand side contour shows the velocity magnitude normalized with the (terminal) film velocity $v_f$ and the right hand side shows the dimensionless rate of viscous dissipation per unit volume normalized using the inertio-capillary scales, represented on a $\log_{\text{10}}$ scale to differentiate the regions of maximum dissipation. Here, the film Ohnesorge number is $\Ohf = 0.10$ and that of air is $\Oha = 10^{-3}$. Also see supplementary movie SM3.}
	\label{fig:three-phaseTemporal}
\end{figure}

Figures~\ref{fig:two-phaseTemporal} and~\ref{fig:three-phaseTemporal} elucidate the two-phase and three-phase Taylor-Culick retractions. For low viscous surroundings ($\Ohs \le 1$), figures~\ref{fig:two-phaseTemporal}a and~\ref{fig:three-phaseTemporal}a show the growth of the dimensionless hole radius ($\tilde{R}(t) = R(t)/h_0$) in time (normalized with the inertio-capillary timescale $\tau_\gamma$), and the insets contain the growth rate of this hole: $\dot{\tilde{R}}_\gamma (t) = \tau_\gamma d\tilde{R}(t)/dt$. After the initial transients, the hole grows (i.e., the film retracts) linearly in time with a constant velocity~($v_f$). We can use this retraction velocity to calculate the film Weber number, 

\begin{align}
	\label{Eqn::Weber definition}
	\Wef \equiv \frac{\rho_fv_f^2h_0}{2\gammasf} = \lim\limits_{\tilde{R} \to \infty}\dot{\tilde{R}}_\gamma^2,
\end{align}

\noindent which is represented by the black dotted lines in figures~\ref{fig:two-phaseTemporal}a and~\ref{fig:three-phaseTemporal}a. $\Wef$ is an output parameter of the retraction process.{Note that, for very low $\Ohs$, as the rim grows with time, the inertial drag on the moving rim due to the surrounding medium overcomes the driving capillary forces, resulting in a decrease of the tip velocity \citep[see insets of figure~\ref{fig:two-phaseTemporal}a and][]{jian_deng_thoraval_2020}. However, we can still calculate a velocity scale (and hence $\Wef$) associated with the Taylor-Culick like retraction immediately after the initial transients (as marked by the black dotted lines in the insets of figure~\ref{fig:two-phaseTemporal}a for the lowest $\Ohs$). 

Furthermore, when the surroundings is highly viscous ($\Ohs \ge 1$), we plot the growing hole radius $\tilde{R}(t)$ as a function of time, which is normalized by the visco-capillary timescale $\tau_\eta$, see \S~\ref{sec::ViscousScaling}, and figures~\ref{fig:two-phaseTemporal}b and~\ref{fig:three-phaseTemporal}b. The insets of these panels contain the growth rate of the hole, calculated as $\dot{\tilde{R}}_\eta = \tau_\eta d\tilde{R}/dt$. Once again, we observe that the growth of the hole (and the film retraction) depend linearly on time with a constant velocity, which can be used to calculate the surroundings capillary number,

\begin{align}
	\label{Eqn::Cas definition}
	\Cas \equiv \frac{\eta_sv_s}{2\gammasf} = \lim\limits_{\tilde{R} \to \infty}\dot{\tilde{R}}_\eta,
\end{align}

\noindent marked with the black dotted lines in figures~\ref{fig:two-phaseTemporal}b,~\ref{fig:three-phaseTemporal}b, and the corresponding insets. $\Cas$ is another output parameter of the retraction process. Note that the velocity of the retracting film ($v_f$) is the same as the velocity scale in the surrounding medium ($v_s$), following the kinematic boundary condition at the circumference of the growing hole. Consequently, the two output parameters, $\Wef$ (\ref{Eqn::Weber definition}) and $\Cas$ (\ref{Eqn::Cas definition}) are related as $\Cas = \Ohs\sqrt{\Wef}$ (see \S~\ref{sec:scaling}). 

Lastly, figures~\ref{fig:two-phaseTemporal}c and~\ref{fig:three-phaseTemporal}c illustrate the flow morphologies for the two-phase and three-phase configurations, respectively, when the hole has grown to $\tilde{R} = 30$. Readers can refer to supplementary movies SM2 and SM3 for temporal dynamics of the two-phase and three-phase configurations, respectively. Similar to the classical Taylor-Culick retraction case (appendix~\ref{App::ClassicalTC} and supplementary movie SM4), both the film and the surroundings move. However, unlike the classical case, even for low $\Ohs$, the surrounding medium takes away momentum from the film owing to inertia (added mass-like effect), thus reducing the retraction velocity (see insets of figures~\ref{fig:two-phaseTemporal}a and~\ref{fig:three-phaseTemporal}a). Furthermore, contrary to the classical case where the dissipation is highest at the neck connecting the rim to the rest of the film (see appendix~\ref{App::ClassicalTC}), the dissipation in the other two configurations is spread out, and also occurs in the surrounding medium. 

As the hole grows, the retracting film collects fluid parcels from upstream of the moving front and forms a rim \citep{culick-1960-japplphys, de1996introductory, villermaux2020fragmentation}. Essentially, the moving fluid parcels of the retracting tip collide with the fluid parcels upstream of the tip, which were initially at rest. The collisions are inelastic as a fraction of the available energy is dissipated by internal viscous fluid friction. For the classical and two-phase configurations, this rim entails a top-bottom symmetry, which is lost in the three-phase configuration. This is due to the air medium having significantly less inertia (added mass-like effect from the properties of the film) than the oil bath, causing the film to dig into the bath and forming a hook-shaped rim (see figures~\ref{fig:two-phaseTemporal}c: i-iii and~\ref{fig:three-phaseTemporal}c: i-iii). Furthermore, the surrounding bath ($s$) engulfs the retracting film in order to feed the precursor film. This is a result of the high capillary pressure (high curvature) in the wedge region near the apparent three-phase contact line \citep{sanjay2019droplet, cuttle2021engulfment}, which also aids in the formation of the hook-shaped rim \citep{peschka-2018-scirep}. Moreover, for the cases where there is a density contrast between the film and the surroundings, the top-bottom symmetry can break down even for the two-phase configuration due to a flapping instability at very low $\Ohs$, as discussed by \citet{lhuissier-2009-prl, jian_deng_thoraval_2020}. Furthermore, as $\Ohs$ increases, the bulbous rim disappears, leading to slender, more elongated retracting films. In the two-phase case, the retraction film maintains (top-bottom) symmetry (see figures~\ref{fig:two-phaseTemporal}c:~iii-v), and the dissipation is highest in the viscous boundary layer in the surrounding medium (see figure~\ref{fig:two-phaseTemporal}c; further elaborated upon in \S~\ref{sec:energetics viscous}). However, the three-phase case features (top-bottom) asymmetric films owing to the accumulation of fluid towards the low-resistance air medium (see figures~\ref{fig:three-phaseTemporal}c: iii-v), and the dissipation is highest near the apparent three-phase contact line (see figure~\ref{fig:three-phaseTemporal}c; further elaborated upon in \S~\ref{sec:energetics viscous}).  The disappearance of bulbous rims matches with the experimental observations \citep[see \S~\ref{sec:3-phaseTC Exp results} and][]{reyssat-2006-epl}. 

Note that the numerical results presented in this section is complementary to the experiments on a film retracting in a submerged oil bath \citep[two-phase case,][]{reyssat-2006-epl} and a film bursting at an air-liquid interface (\S~\ref{sec:3-phaseTC Exp results}). The numerical simulations give us access to the cross-sectional view to elucidate the shape of the retracting films (figures~\ref{fig:two-phaseTemporal} and~\ref{fig:three-phaseTemporal}), which is difficult to resolve experimentally. On the other hand, our axisymmetric (by definition) simulations do not show the azimuthal instabilities resulting in the corrugations at the oil-air-water contact line. Furthermore, as we focus only on the early time dynamics, we also neglect the curvature of the oil drop in the case of the three-phase retractions. Nonetheless, we can still sufficiently compare the dependences of the retraction velocity on the Ohnesorge number $\Ohs$ of the surroundings (see \S~\ref{sec:scaling}), along with the scaling relations that we develop in the next section for both the experimental and numerical datapoints. 

\section{Taylor-Culick retractions: a force perspective}\label{sec:forces}

The capillary and viscous forces, along with the inertia of the film and the surrounding media, govern the retraction dynamics. For the classical configuration (figure~\ref{fig:configs}a), the viscosity and inertia of the outer medium are negligible. Furthermore, the film viscosity~$\eta_f$ plays no role in determining the magnitude of the retraction velocity owing to the internal nature of the associated viscous stresses \citep{savva-2009-jfm}, as long as $Oh_{f}$ is less than the aspect ratio of the film \citep[see][]{deka-2020-prf}. Using these features, \citet{taylor-1959-procrsoclonda} calculated $v_f = v_{\text{TC}}$ (\ref{eq:v_TC}), resulting solely from momentum equilibrium while disregarding the fate of the liquid accumulated in the rim \citep{villermaux2020fragmentation}. In terms of the dimensionless numbers introduced earlier (see \S~\ref{sec:Num results}), (\ref{eq:v_TC}) implies that $\Wef = \rho_fv_f^2h_0/(2\gammaaf)$ is constant and equal to 1 (see appendix~\ref{App::ClassicalTC} for details of the retraction dynamics in the classical configuration). In this section, we delve into the different realizations of the dominating forces, and their implications, in the two-phase and three-phase configurations.  

\subsection{Two-phase Taylor-Culick retractions}\label{sec:2-phase forces}
For the two-phase configuration (figure~\ref{fig:configs}b), if the viscosity of the oil phase is small (i.e., $\Ohs = \eta_s/\sqrt{\rho_f\gammasf h_0} \ll 1$), the Weber number based on the film velocity $v_f$ and the driving surface tension coefficient ($2\gammasf$), $\Wef = \rho_fv_f^2h_0/(2\gammasf)$ (\ref{Eqn::Weber definition}) has a value smaller than 1 (see inset of figure~\ref{fig:two-phaseTemporal}a). Nonetheless, the driving surface tension force $F_\gamma(t) \sim \gammasf\left(2\pi R(t)\right)$ (see figure~\ref{fig:configs}b) still balances the inertial force $F_\rho(t) \sim \rho_f v_f^2\left(2\pi R(t)\right)h_0$. Note that since the oil (surrounding) and water (film) densities are very similar ($\rho_f \approx \rho_s$), we can still use $\rho_f$ for the density scale despite the added mass-like effect. Consequently, in this regime, the Weber number is still a constant during retraction ($\Wef \sim \mathcal{O}\left(1\right)$,  inset of figure~\ref{fig:two-phaseTemporal}a). 

On the other hand, if the viscosity of the oil phase ($\eta_s$) is significantly higher (i.e., $\Ohs \gg 1$), the resistive viscous force $F_{\eta}(t)$ dominates over the inertial effects, as the surroundings Reynolds number $Re_s \equiv \rho_s v_s h_0/\eta_s \sim \mathcal{O}$($10^{-2}$). In such a scenario, the retraction dynamics will be governed by the balance between the capillary ($F_{\gamma}(t)$) and viscous ($F_{\eta}(t)$) forces \citep{fraaije-1989-jcis, reddy-2020-prf}, given by

\begin{align}
F_{\gamma}(t) \sim F_{\eta}(t),
\label{eq:2 phase balance 1}
\end{align}

\noindent where (from figure~\ref{fig:configs}b)

\begin{align}
F_\gamma(t) = 2 \gamma_{sf}\left(2\pi R(t)\right).
\label{eq:2 phase gamma}
\end{align}

For $F_{\eta}(t)$ in (\ref{eq:2 phase balance 1}), one can consider the retracting rim to be a cylinder translating in a viscous flow \citep{reyssat-2006-epl, eri-2010-pre}. Thus, the viscous drag can be described by the Oseen approximation to the Stokes flow \citep{book-lamb, book-happel}, which to the leading order is expressed as

\begin{align}
F_{\eta}(t) \sim \eta_{s} v_f\left(2\pi R(t)\right).
\label{eq:2 phase viscous force}
\end{align}

\noindent where the factor $2\pi R(t)$ is due to the axisymmetric geometry. On equating (\ref{eq:2 phase gamma}) and~(\ref{eq:2 phase viscous force}), we get

\begin{align}
v_f \sim \frac{\gammasf}{\eta_{s}} .
\label{eq:2 phase v}
\end{align}

Moreover, $v_f = v_s$ (where $v_s$ is the velocity scale in the surrounding medium, see \S~\ref{sec:Num results}). As a result, (\ref{eq:2 phase v}) implies that the capillary number $\Cas$ (\ref{Eqn::Cas definition}) is constant, i.e.,
 
\begin{align}
	\Cas = \frac{\eta_s v_s}{2\gammasf} \sim \mathcal{O}\left(1\right) .
	\label{eq:2 phase Ca}
\end{align}

Further, upon dividing both sides of (\ref{eq:2 phase v}) by the inertio-capillary velocity scale $v_\gamma = \sqrt{2\gammasf/(\rho_fh_0)}$ and squaring, we obtain

\begin{align}
\Wef \sim \Ohs^{-2} .
\label{eq:2 phase We}
\end{align}

\noindent The aforementioned equations (\ref{eq:2 phase Ca}) -- (\ref{eq:2 phase We}) denote the scaling laws for viscous two-phase Taylor-Culick retractions. 

\subsection{Three-phase Taylor-Culick retractions}\label{sec:3-phase forces}
For the three-phase configuration (figure~\ref{fig:configs}c), in the viscous limit ($\Ohs \gg 1$), the force balance is still given by (\ref{eq:2 phase balance 1}), especially for the oils that are significantly more viscous than water. Here, the driving surface tension force can be expressed as (from figures~\ref{fig:configs}c and~\ref{fig:NumericalSetup}b)

\begin{align}
F_\gamma(t) = (\gammasf + \gammasa + \gammasf - \gammasa) \left(2\pi R(t)\right)= 2 \gammasf\left(2\pi R(t)\right),
\label{eq:3 phase gamma}
\end{align}

\noindent assuming the presence of a precursor film of oil on top of the water film \citep[see \S~\ref{sec:3-phaseTC Num methods} and][]{book-degennes, bonn2009wetting, thoraval-2013-pre}. However, writing an expression for $F_{\eta}(t)$ is not as straightforward as the two-phase configuration. As can be observed from figure~\ref{fig:three-phaseTemporal}c, during the retraction of the film, the oil climbs on top of the water, resulting in a strong flow in the wedge-like region close to the oil-air-water contact line. The rate of local viscous dissipation in this region is also very high (right panels of figure~\ref{fig:three-phaseTemporal}c). Similar wedge flows have also been observed for moving contact lines on solid substrates \citep{degennes-1985-rmp, marchand-2012-prl, snoeijer-2013-arfm}. It has been reported that the wedge flow results in a viscosity-dependence of velocity that is weaker than 1/$\eta_{s}$ \citep{marchand-2012-prl}, but the exact nature of the dependence has hitherto not been quantified. The presence of a deformable liquid substrate on which the wedge flow occurs (the retracting water film in this case) complicates the situation even further -- making it extremely difficult to arrive at the experimentally-observed $v_f(\eta_{s}) \sim \eta_{s}^{-1/2}$ dependence (\ref{eq:exp scaling}) from a simple force balance. In \S~\ref{sec:energetics viscous}, we attempt to explain this scaling from an energetics point of view. Nevertheless, from the experiments, we know that the $v_f (\eta_{s})$ scaling is given by (\ref{eq:exp scaling}), which can be rewritten as

\begin{align}
\Cas \sim \Ohs^{1/2} .
\label{eq:3 phase Ca}
\end{align}
Dividing both sides of (\ref{eq:3 phase Ca}) by $v_\gamma$ from (\ref{Eqn::Scales}) and squaring, we obtain
\begin{align}
\Wef \sim \Ohs^{-1} .
\label{eq:3 phase We}
\end{align}

Therefore, from (\ref{eq:2 phase v}), (\ref{eq:2 phase We}), (\ref{eq:3 phase Ca}), and (\ref{eq:3 phase We}), we hypothesize that the presence of the oil-air-water apparent contact line in the three-phase configuration dramatically alters the scaling relationships as compared to the two-phase configuration for $\Ohs > 1$ (see figures~\ref{fig:two-phaseTemporal}b and~\ref{fig:three-phaseTemporal}b). This will be further elaborated upon in \S~\ref{sec:energetics viscous}. Contrary to this scenario, for low $\Ohs$ numbers, the retraction velocities in both these configurations have the same scaling behavior. Despite the presence of a hook-shaped rim in the three-phase case (figure~\ref{fig:three-phaseTemporal}c: i-ii), we can still treat the moving rim and the surroundings as lumped elements. As a result, the driving surface tension force $\gammasf\left(2\pi R(t)\right)$ still balances the inertial force that scales with $\rho_f v_f^2\left(2\pi R(t)\right)h_0$, thus giving $\Wef \sim \mathcal{O}\left(1\right)$ (see figure~\ref{fig:three-phaseTemporal}a). In the next section, we demonstrate the validity of the scaling relations developed in this section.

\section{Demonstration of the scaling relationships}\label{sec:scaling}
\begin{figure}
	\centering
	\includegraphics[width=\textwidth]{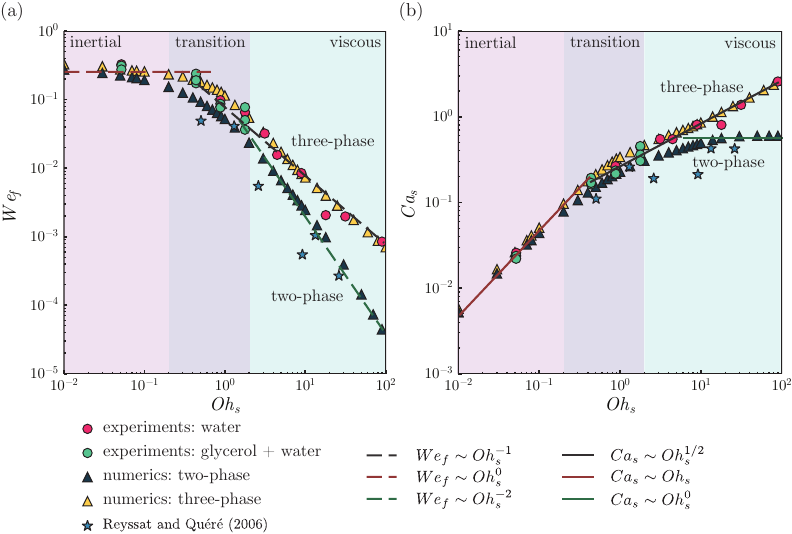}
	\caption{Regime maps visualized as $\Wef$ vs. $\Ohs$ in (a) and as $\Cas$ vs. $\Ohs$ in~(b). The experimental datapoints (circles) correspond to the three-phase configuration (figure~\ref{fig:configs}c) while the simulations (triangles) correspond to both the two-phase (figure \ref{fig:configs}b) and three-phase (figure~\ref{fig:configs}c) configurations. The experimental datapoints (pentagrams) for the two-phase configuration have been adopted from \citet{reyssat-2006-epl} for their silicone oil~(surroundings,~$s$) -- soap water (film, $f$) dataset.}
	\label{fig:scalingFinal}
\end{figure}

Figure~\ref{fig:scalingFinal} illustrates the dependence of $\Wef$ and $\Cas$ on the Ohnesorge number $\Ohs$ of the surroundings for the retraction configurations described in figure~\ref{fig:configs}. Note that the same datapoints are presented in both panels~\ref{fig:scalingFinal}a and~\ref{fig:scalingFinal}b, following the relation $\Cas~=~\Ohs\sqrt{\Wef}$ (as $v_f = v_s$, see \S~\ref{sec:Num results}). In figure~\ref{fig:scalingFinal}a, $\Wef = 1$ marks the classical Taylor-Culick retraction limit, whereas for the two-phase and three-phase configurations, we identify two regimes: inertial ($\Ohs < 1$) and viscous ($\Ohs > 1$). 

The inertial scaling is identical for both the two-phase and three-phase configurations: $\Wef \sim \mathcal{O}\left(1\right)$, which also implies $\Cas \sim \Ohs$ (see \S~\ref{sec:2-phase forces} and~\ref{sec:3-phase forces}). The brown lines in figure~\ref{fig:scalingFinal} represent these two scaling relations.

The datapoints corresponding to the two-phase numerical simulations (from figure~\ref{fig:two-phaseTemporal}) are shown by the dark blue triangles. As $\Ohs$ increases, the retraction transitions from the inertial scaling (brown lines), to the viscous two-phase Taylor-Culick scaling:~$\Cas~\sim~\Ohs^0$ (\ref{eq:2 phase Ca}) or $\Wef \sim \Ohs^{-2}$ (\ref{eq:2 phase We}). We also plot the experimental datapoints from \citet{reyssat-2006-epl} for their silicone oil (surroundings, $s$) -- soap water (film, $f$) dataset, shown in figure~\ref{fig:scalingFinal} by the light blue pentagrams. In order to make these datapoints dimensionless, we use $h_0 = 100\,\si{\micro\meter}$ and $\gammasf = 7\,\si{\milli\newton}/\si{\meter}$, denoting the thickness of the soap film and the surroundings-film interfacial tension coefficient, respectively, in their experiments. We also neglect any Marangoni flow, or dynamic surface tension effects. Our simulations and scaling relationships are in reasonable agreement with the experimental datapoints of \citet{reyssat-2006-epl}. Note that \citet{reyssat-2006-epl} tried to fit a trend line of $\left(\ln\eta_{s}\right)/\eta_{s}$ (higher order Oseen correction) through all of their experimental datapoints to obtain a good fit. When the same datapoints are plotted in figure~\ref{fig:scalingFinal}a and b, it is observed that some of their datapoints (corresponding to the low $\Ohs$ numbers) are, in fact, in the transition between the inertial and the viscous regimes, while the rest of the datapoints show reasonable agreement with the viscous two-phase retraction dynamics given by (\ref{eq:2 phase Ca}) or (\ref{eq:2 phase We}). 

We also plot the datapoints corresponding to our experiments (figure~\ref{fig:dynamics}) and simulations (figure~\ref{fig:three-phaseTemporal}) for the three-phase configuration (figure~\ref{fig:configs}c) in figure~\ref{fig:scalingFinal}. We observe that, at low $\Ohs$, these datapoints follow the inertial dynamics (brown lines), while at higher $\Ohs$, the datapoints follow the scaling relationships given by (\ref{eq:3 phase Ca}) and (\ref{eq:3 phase We}): $\Cas \sim \Ohs^{1/2}$ and $\Wef \sim \Ohs^{-1}$, respectively (represented by the black lines in figure~\ref{fig:scalingFinal}). Note that in order to non-dimensionalize the experimental datapoints shown in figure~\ref{fig:dynamics}c (so that they can be plotted in figure~\ref{fig:scalingFinal}), one needs to know the film thickness $h_0$. In the present experiments, the optical resolution was insufficient for accurate measurement of the film thickness prior to rupture. Moreover, as mentioned earlier, the breakup process itself is highly sensitive to experimental noise \citep[see \S~4.2 of][]{villermaux2020fragmentation}. Similar difficulties were also presumably experienced by \citet{eri-2010-pre} in their experiments of two-phase retraction, and they used a fitting parameter in their $v_f (\eta_{s})$ relation, which was a function of $h_0$. We also know from bubble bursting experiments \citep{doubliez-1991-ijmf, lhuissier-2012-jfm} that the film thickness prior to breakup varies in the range $\mathcal{O}\left(100\,\si{\nano\meter}\right)$ -- $\mathcal{O}\left(10\,\si{\micro\meter}\right)$. Moreover, in similar studies \citep{lhuissier-2012-jfm, thoroddsen-2012-jfm}, the film thickness is retroactively calculated from the retraction velocity measurements. We can see from figure~\ref{fig:scalingFinal} that for low $\Ohs$, the dynamics are independent of the specific nature of the configuration (classical, two-phase, or three-phase). Knowing $v_f$, $\eta_{s}$, and $\gammasf$, we can calculate the $\Cas$ for the datapoint in figure~\ref{fig:dynamics}c corresponding to $\eta_{s}$ = $4.94 \times 10^{-4}\,\si{\pascal}.\si{\second}$. Fitting that $\Cas$ value to the $\Cas \sim \Ohs$ trend line (brown line) in figure~\ref{fig:scalingFinal}b, a value of $h_0 = 1.5\,\si{\micro\meter}$ can be calculated, which is within the range observed for previous experiments in a similar system \citep{lhuissier-2012-jfm}. Using $h_0 = 1.5\,\si{\micro\meter}$ for the remaining experimental datapoints in figure~\ref{fig:dynamics}c (for $\eta_{s} >$ 4 $\times$ 10$^{-3}$ Pa.s) and calculating $\Cas$, $\Ohs$, and $\Wef$, we find that those datapoints (red circles) also collapse on the trend lines (black lines) along with the numerical simulations (yellow triangles) in figure~\ref{fig:scalingFinal}. Note that a water film thickness of $h_0 = 1.5\,\si{\micro\meter}$ sets the $\Ohf$ at $0.1$ for the three-phase case, which is different from the $\Ohf$ that we use for the two-phase case \citep[$\Ohf = 0.05$ based on their experiments of][]{reyssat-2006-epl}. Therefore, to justify comparison between the two cases, we varied $\Ohf$ in simulations from $0.01$ to $0.1$ and found that the dimensionless retraction velocities ($\Wef$ and $\Cas$) are $\Ohf$-independent for both the two-phase and three-phase configurations (for $\Ohf < 1$). We also verify the $\Ohf$-independence experimentally by replacing the water in our bath by glycerol-water mixtures, and the measurements thus obtained (green circles) also follow the $\Wef \sim \Ohs^{-1}$ and $\Cas \sim \Ohs^{1/2}$ trendlines (black lines) in figures~\ref{fig:scalingFinal}a and~\ref{fig:scalingFinal}b, respectively.  

In summary, in \S~\ref{sec:forces} -- \S~\ref{sec:scaling}, we discussed the forces involved during the retraction of liquid films owing to the unbalanced capillary traction, followed by identification of the inertial ($\Ohs < 1$) and viscous ($\Ohs > 1$) regimes in the $\Wef$ vs. $\Ohs$ and $\Cas$ vs. $\Ohs$ dependences. We also checked the validity of the corresponding scaling behaviors in this section. To further understand the retraction dynamics due to the capillary traction, we focus on the different thermodynamically consistent energy transfer modes in the next section. Particularly, we try to understand the scaling relationship for the viscous three-phase Taylor-Culick retraction that still eludes understanding from a momentum balance point of view (see \S~\ref{sec:energetics viscous}). 

\section{Taylor-Culick retractions: an energetics perspective} \label{sec:energetics}
\begin{figure}
	\centering
	\includegraphics[width=\textwidth]{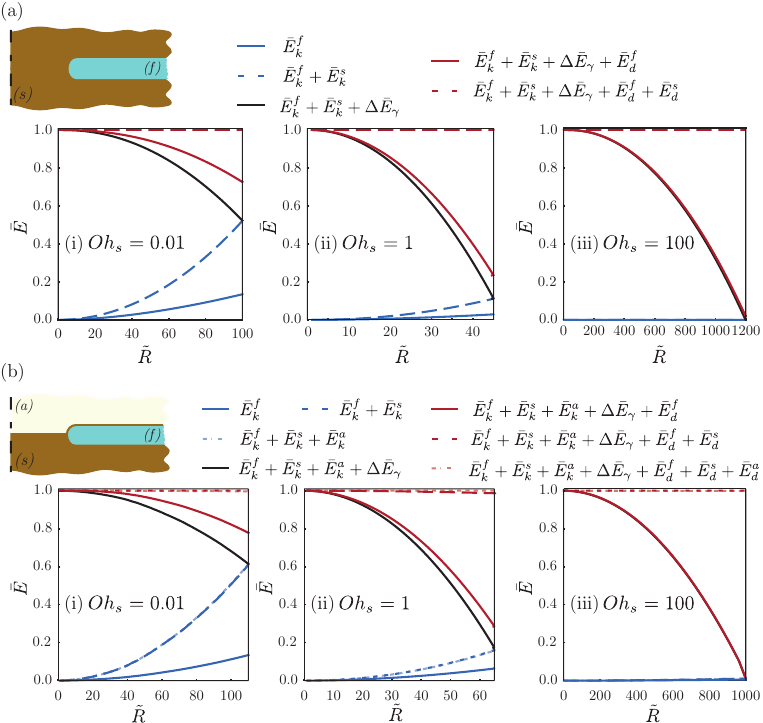}
	\caption{Energy budget at different $\Ohs$ for the (a) two-phase and (b) three-phase configurations. The energies $E$ are normalized by the total surface energy released as the film retracts creating a hole of radius $\tilde{R}_{\text{max}} = 100$ for $Oh_s \le 1$, and $\tilde{R}_{\text{max}} = 1000$ (two-phase case) and $\tilde{R}_{\text{max}} = 1200$ (three-phase case) for $Oh_s = 100$. Note that this $\tilde{R}_{\text{max}}$, and hence the surface energy datum, are arbitrarily chosen. We use hole radii that are large enough such that the sheets approach a constant velocity. The superscripts account for the film ($f$), the surroundings ($s$), and air ($a$). }
	\label{fig:energy}
\end{figure}

\begin{figure}
	\centering
	\includegraphics[width=\textwidth]{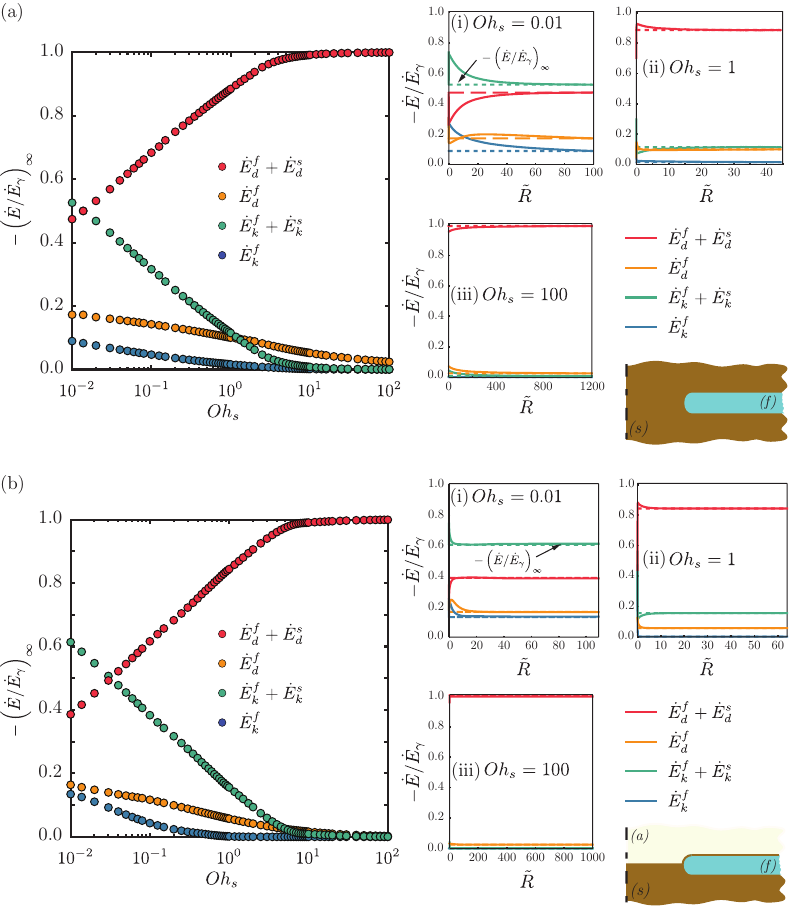}
	\caption{Variation of the rate of change of kinetic energy ($\dot{E}_k$) and viscous dissipation ($\dot{E}_d$) as proportions of the rate of energy injection ($-\dot{E}_\gamma$) with $\Ohs$ at steady state for the (a) two-phase and (b) three-phase configurations. For both cases, in the inertial limit ($\Ohs \ll 1$), the fraction of energy that goes into kinetic energy and viscous dissipation are comparable. However, in the viscous limit ($\Ohs \gg 1$), viscous dissipation in the surroundings dominates. Insets show the representative temporal variations of the ratio of the rate of change of energy ($\dot{E}$) to the rate of energy injection ($-\dot{E}_{\gamma}$) with dimensionless hole radius $\tilde{R}$ at three different $\Ohs$.  The superscripts account for the film ($f$), the surroundings ($s$), and air ($a$). }
	\label{fig:energy_rate}
\end{figure}

A retracting liquid sheet loses surface area and consequently releases energy \citep{dupre1867theorie, dupre1869theorie, rayleigh-1891-nature, culick-1960-japplphys}, which further increases the kinetic energy of the system (i.e., film and surroundings). A part of this energy is lost in the process due to viscous dissipation. So, the overall energy budget entails

\begin{align}
	E_k^f(R) + E_k^s(R) + E_k^a(R) + E_\gamma(R) + E_d^f(R) + E_d^s(R) + E_d^a(R) = E_\gamma(R = 0),
	\label{Eqn::OverallEnergyBalance}
\end{align}

\noindent where $E_\gamma$ is the surface energy, $E_k$ the kinetic energy, and $E_d$ the viscous dissipation. The superscripts account for the film ($f$), the surroundings ($s$), and air ($a$). Of course, for the two-phase case, the terms associated with air ($a$) do not exist as there is no air phase. Figure~\ref{fig:energy} depicts (\ref{Eqn::OverallEnergyBalance}) for both the two-phase and three-phase configurations. Coincidentally, even for the three-phase configuration, the energies associated with the air phase are negligible (see figure~\ref{fig:energy}, the dashed and dot-dashed lines overlap), even though the velocity field in air is not negligible (figure~\ref{fig:three-phaseTemporal}c). We keep $E^a_k(R)$ and $E^a_d(R)$ in the energy budget for the sake of completeness. In general, the sum of all these energies at any hole radius $R(t)$ equals the total surface energy at $R = 0$, i.e., the total energy available to the system. As the film retracts, it continuously releases energy, as its surface energy decreases. Therefore, to calculate (\ref{Eqn::OverallEnergyBalance}), one can choose a reference for surface energy arbitrarily. In figure~\ref{fig:energy}, the surface energy at a hole radius of $R = R_{\text{max}}$ is used as this arbitrary instance. This datum is chosen such that by the time the hole expands to $R_{\text{max}}$, the film would have reached a constant velocity. Furthermore, we can normalize the energies in (\ref{Eqn::OverallEnergyBalance}) with the total surface energy released as the film retracts to a hole of radius $R_{\text{max}}$. The energy budget now reads 

\begin{align}
	\bar{E}_k^f(\tilde{R}) + \bar{E}_k^s(\tilde{R}) + \bar{E}_k^a(\tilde{R}) + \Delta \bar{E}_\gamma(\tilde{R}) + \bar{E}_d^f(\tilde{R}) + \bar{E}_d^s(\tilde{R}) + \bar{E}_d^a(\tilde{R}) = 1.
	\label{Eqn::OverallEnergyBalance2}
\end{align}

\noindent Here, $\bar{E}(\tilde{R}) = E(\tilde{R})/(E_\gamma(0)-E_\gamma(\tilde{R}_{\text{max}}))$, $\Delta E_{\gamma}(\tilde{R}) = E_{\gamma}(\tilde{R}) - E_{\gamma}(\tilde{R}_{\text{max}})$, and $\tilde{R}~=~\tilde{R}(t)~=~R(t)/h_{0}$ is the dimensionless hole radius. The reader is referred to appendix~\ref{App::EnergyBalance} for details of the energy budget calculations. 

Removing the arbitrary datum described above and noting that there is a continuous injection of surface energy ($-\dot{E}_\gamma$, minus sign because the surface energy is decreasing with the growing hole) into the system, we can also write the energy budgets in terms of rates:

\begin{align}
		\label{Eqn::EnergyRates2}
		\dot{E}_k^f(\tilde{R}) + \dot{E}_k^s(\tilde{R}) + \dot{E}_d^f(\tilde{R}) + \dot{E}_d^s(\tilde{R}) = \left(-\dot{E}_\gamma(\tilde{R})\right).
\end{align}

Figure~\ref{fig:energy_rate} visualizes (\ref{Eqn::EnergyRates2}) by plotting the proportion of the rate of surface energy released that goes into the rate of increase of kinetic energy and the rate of total viscous dissipation. From the insets (i -- iii) of this figure, we observe that these fractions saturate after initial transients. So, we also plot these steady state values (\ref{Eqn::EnergyRatesPlotting}) in panels~\ref{fig:energy_rate}a and~\ref{fig:energy_rate}b for the two-phase and three-phase configurations, respectively, 

\begin{align}
	\left(\dot{E}/\dot{E}_{\gamma}\right)_\infty = \lim\limits_{\tilde{R} \to \infty}\left(\frac{\dot{E}(\tilde{R})}{\dot{E}_{\gamma}(\tilde{R})}\right).
	\label{Eqn::EnergyRatesPlotting}
\end{align}

We devote the rest of this paper to understanding the distribution of the energy injection rate into the rates of increase of kinetic energy and viscous dissipation for both the inertial and viscous regimes. 

\subsection{Energy transfers in the inertial regime}\label{sec:energetics inertial}
We first focus on the energy balance in the classical Taylor-Culick retraction and the famous  Dupr{\'e}-Rayleigh paradox \citep{villermaux2020fragmentation}. \citet{dupre1867theorie, dupre1869theorie} hypothesized that the total surface energy released during retraction manifests as the kinetic energy of the film \citep{rayleigh-1891-nature}. As a result, the predicted retraction velocity was off by a factor of $\sqrt{2}$ (see appendix~\ref{App::ClassicalTC}), leading to discrepancies with experiments \citep{ranz-1959-japplphys, culick-1960-japplphys}. Nonetheless, it is noteworthy that \citet{dupre1867theorie, dupre1869theorie} reached the correct scaling relationship by identifying the essential governing parameters of classical sheet retractions. 

\citet{culick-1960-japplphys} identified that the rate of surface energy released (\ref{Eqn::Surface Energy rate}) should be distributed into an increase in kinetic energy of the rim and the viscous dissipation inside the film: $-\dot{E}_\gamma(t) = \dot{E}_k^f(t) + \dot{E}_d^f(t)$. The viscous dissipation can be attributed to the inelastic acceleration of the undisturbed film up to the velocity of the edge of the rim. Note that the dissipation is independent of the fluid viscosity and is given by \citep{culick-1960-japplphys}

\begin{align}
	\dot{E}_d^f(t) = \frac{1}{2}\frac{dm(t)}{dt}v_f^2 , 
	\label{Eqn::Classical Dissipation}
\end{align}
where $m(t)$ is the mass of the retracting film.

Coincidentally, this rate of viscous dissipation in the film is the same as the rate of increase in its kinetic energy \citep[constant rim velocity,][]{culick-1960-japplphys, villermaux2020fragmentation}. We confirm this hypothesis in appendix~\ref{App::ClassicalTC} \citep[see figures~\ref{fig:ClassicalTC}c, d, and][]{sunderhauf-2002-pof}, whereby

\begin{align}
	\dot{E}_k^f(t) \approx \dot{E}_d^f(t) \approx - \dot{E}_\gamma(t)/2.
\end{align}

Next, we delve into the energy transfers in the two-phase and three-phase configurations. In the inertial limit, in a manner akin to the classical case, the fraction of the rate of energy injection that goes into increasing the kinetic energy is similar to that of viscous dissipation. However, unlike the classical case, the kinetic energy as well as viscous dissipation are distributed among the film and the surrounding medium (figures~\ref{fig:energy} and~\ref{fig:energy_rate}, $\Ohs \ll 1$). We observe that

\begin{align}
	\left(\dot{E}_d^f(t) + \dot{E}_d^s(t)\right) \approx \left(\dot{E}_k^f(t) + \dot{E}_k^s(t)\right) \approx -\dot{E}_\gamma(t)/2.
\end{align}

\noindent In a manner reminiscent of \citet{dupre1867theorie, dupre1869theorie}, we can write

\begin{align}
	-\dot{E}_\gamma(t) \approx \left(\dot{E}_k^f(t) + \dot{E}_k^s(t)\right) \sim \left(\rho_f v_fh_0\left(2\pi R(t)\right)\right)v_f^2,
	\label{Eqn::ReminiscentDupre}
\end{align}

\noindent where $v_f = v_s$ (kinematic boundary condition at the tip of the film) and $\rho_s = \rho_f$. Additionally, following \citet{bohr2021surface} and appendix~\ref{App::EnergyBalance}, the rate of change of surface energy is given by

\begin{align}
	\dot{E}_\gamma(t) \approx -F_{\gamma}(t) \frac{dR(t)}{dt} = -2\gammasf\left(2\pi R(t)\right)v_f.
	\label{Eqn::Surface Energy rate}
\end{align}

\noindent Using (\ref{Eqn::ReminiscentDupre}) -- (\ref{Eqn::Surface Energy rate}), and rearranging, we get

\begin{align}
		\Wef = \frac{\rho_fv_f^2h_0}{2\gammasf} \sim \mathcal{O}\left(1\right),
\end{align}

\noindent which is the same as the inertial scaling derived using the force balance (insets of figures \ref{fig:two-phaseTemporal}a and \ref{fig:three-phaseTemporal}a).

\subsection{Demystifying dissipation in the viscous regime}\label{sec:energetics viscous}
\begin{figure}
	\centering
	\includegraphics[width=\textwidth]{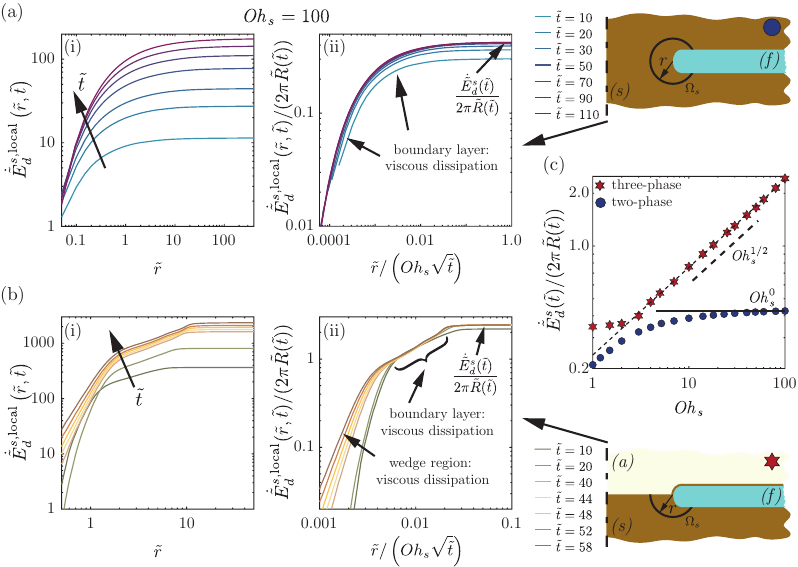}
	\caption{Dissipation in the viscous limit ($\Ohs \gg 1$) of Taylor-Culick retractions: evolution of the local rate of viscous dissipation $\left(\dot{\tilde{E}}_d^{s, \text{local}}\left(\tilde{r}, \tilde{t}\right)\right)$ with dimensionless distance $\tilde{r} = r/h_0$ away from (a) the tip of the film in the two-phase configuration and (b) the macroscopic three-phase contact line in the three-phase configuration. In insets (ii), this distance is normalized with the dimensionless viscous boundary layer thickness in the surrounding medium, $\tilde{\delta}_\nu = \delta_\nu/h_0 = \Ohs\sqrt{\tilde{t}}$. Here, $\tilde{R} = R/h_0$ and $\tilde{t} = t/\tau_\eta$ are the dimensionless hole radius and dimensionless time, respectively. (c) Variation of the total viscous dissipation rate per unit circumference of the hole $\left(\dot{\tilde{E}}_d^{s}\left(\tilde{t}\right)/\left(2\pi\tilde{R}\left(\tilde{t}\right)\right)\right)$ at steady state with the surroundings Ohnesorge number $\Ohs$.}
	\label{fig:dissipation}
\end{figure}

In the viscous limit $\left(\Ohs \gg 1\right)$, for both the two-phase and three-phase configurations, the surface energy released is entirely dissipated in the surrounding medium (figures~\ref{fig:energy} and~\ref{fig:energy_rate}), i.e.,

\begin{align}
	-\dot{E}_\gamma(t) \sim \dot{E}_d^{s}(t).
	\label{Eqn::Viscous energy balance}
\end{align}

\noindent In fact, this interplay between the surface energy and the viscous dissipation sets the velocity scale $\left(v_s\right)$ in the surrounding medium, which is equal to the retraction velocity ($v_f$, kinematic boundary condition at the hole). Therefore, to estimate this velocity, we first calculate the rate of viscous dissipation $\dot{E}_d^{s}(t)$, which depends on both the viscosity $\eta_{s}$ of the surrounding medium and the velocity gradients $\boldsymbol{\mathcal{D}}$, following the relation (see appendix~\ref{App::EnergyBalance})

\begin{align}
	\dot{E}_d^s = \int_{\Omega_s} 2\eta_{s} \left(\boldsymbol{\mathcal{D}}:\boldsymbol{\mathcal{D}}\right) d\Omega_s = \int_{\Omega_s}\varepsilon_s d\Omega_s.
	\label{Eqn::Viscous dissipation surr}
\end{align}

\noindent Here, $\varepsilon_s$ is the rate of viscous dissipation per unit volume, and the integrals are evaluated over the volume $\Omega_{s}$ of the surrounding medium. Note that $\varepsilon_s$ is highest at the expanding hole, i.e., the tip of the retracting film in the case of two-phase retractions (figures~\ref{fig:two-phaseTemporal}c), and the macroscopic contact line in the case of three-phase retractions (figures~\ref{fig:three-phaseTemporal}c). The latter is analogous to wetting and dewetting of rigid surfaces \citep{degennes-1985-rmp, bonn2009wetting, snoeijer-2013-arfm}. Motivated by this analogy, we calculate the local rate of viscous dissipation integrated over volume elements $\Omega_s(r)$ centered at the expanding hole,

\begin{align}
	\dot{E}_d^{s, \text{local}}(r, t) = \int\limits_{0}^{\Omega_s(r)} \varepsilon_s(r, t) d\Omega_s.
	\label{Eqn::Viscous dissipation surr local}
\end{align}

\noindent where $r$ is the radial distance away from the hole (see insets of figure~\ref{fig:dissipation}c). Additionally, in the viscous regime, we can use the visco-capillary velocity $v_\eta = 2\gammasf/\eta_s$ and the film thickness $h_0$ to non-dimensionalize (\ref{Eqn::Viscous dissipation surr local}) \citep[see \S~\ref{sec::ViscousScaling} and][]{stone1989relaxation}, 

\begin{align}
	\dot{\tilde{E}}_d^{s, \text{local}}\left(\tilde{r}, \tilde{t}\right) \equiv \frac{\dot{E}_d^{s, \text{local}}\left(\tilde{r}, \tilde{t}\right)}{\eta_sv_\eta^2h_0} = \int\limits_{0}^{\tilde{\Omega}_s(\tilde{r})} \tilde{\varepsilon}_s\left(\tilde{r}, \tilde{t}\right) d\tilde{\Omega}_s.
	\label{Eqn::Viscous dissipation surr dimless}
\end{align}

Figures~\ref{fig:dissipation}a-i and~\ref{fig:dissipation}b-i show that the local viscous dissipation increases as we move away from the hole (increasing $\tilde{r}$). Furthermore, the energy dissipated increases in time as the region of flow expands, owing to the increasing hole radius and the dominant radial flow. To rationalize this increase, we plot the rate of local viscous dissipation per unit circumference of the hole in figures~\ref{fig:dissipation}a-ii and~\ref{fig:dissipation}b-ii.

For the two-phase case, the viscous dissipation occurs in the viscous boundary layer $\left( \tilde{\delta}_\nu \sim \Ohs\sqrt{\tilde{t}}\right)$ and saturates at $\tilde{r} \approx \tilde{\delta}_\nu$ (figure~\ref{fig:dissipation}a-ii). However, for the three-phase case, we can identify two distinct regions of viscous dissipation, the wedge region close to the macroscopic contact line, where the viscous dissipation per unit circumference of the expanding hole increases steeply $\left(\tilde{r} < 0.01\tilde{\delta}_\nu\right)$, and the viscous boundary layer $\left(\tilde{r} < 0.1\tilde{\delta}_\nu\right)$, beyond which it saturates (figure~\ref{fig:dissipation}b-ii). Furthermore, this saturation value gives the total viscous dissipation per unit circumference of the hole,

\begin{align}
	\frac{\dot{\tilde{E}}_d^s(\tilde{t})}{\left(2 \pi \tilde{R}(\tilde{t})\right)} = \lim_{\tilde{r} \to \infty}\frac{\dot{\tilde{E}}_d^{s, \text{local}}\left(\tilde{r}, \tilde{t}\right)}{\left(2 \pi \tilde{R}(\tilde{t})\right)},
\end{align}

\noindent which is shown in figure~\ref{fig:dissipation}c as a function of $\Ohs$. We observe that for the two-phase case, the total dissipation is independent of $\Ohs$, whereas in the three-phase case, it scales with $\Ohs^{1/2}$.

\begin{align}
	\label{Eqn::Dissipation Ohs}
	\dot{\tilde{E}}_d^{s}(\tilde{t}) \sim 
	\begin{cases}
			\Ohs^{0}\left(2\pi\tilde{R}(\tilde{t})\right)&\text{two-phase case},\\
			\\
			\Ohs^{1/2}\left(2\pi\tilde{R}(\tilde{t})\right)&\text{three-phase case}.
	\end{cases}
\end{align}

Moreover, upon non-dimensionalizing (\ref{Eqn::Surface Energy rate}) using the same scales as used in (\ref{Eqn::Viscous dissipation surr dimless}), and noting that $v_f = v_s$ and $\Cas = \eta_sv_s/(2\gammasf)$, we get

\begin{align}
	-\dot{\tilde{E}}_\gamma(t) \equiv \frac{\gammasf v_f\left(2 \pi R(t)\right)}{\eta_sv_\eta^2h_0} = \Cas \left(2 \pi \tilde{R}(\tilde{t})\right).
	\label{Eqn::Surface Energy rate dimless}
\end{align} 

\noindent Lastly, equating (\ref{Eqn::Dissipation Ohs}) and (\ref{Eqn::Surface Energy rate dimless}), we get,

\begin{align}
	\label{Eqn::Cas Ohs}
	Ca_s \sim 
	\begin{cases}
		\Ohs^{0} &\text{two-phase case},\\
		\\
		\Ohs^{1/2} &\text{three-phase case}.
	\end{cases}
\end{align}

In summary, in this section, we confirmed our hypothesis that the presence of the oil-air-water contact line in the three-phase configuration dramatically alters the scaling relationships and dynamics as compared to the two-phase configuration (see \S~\ref{sec:3-phase forces}). We also relate the dimensionless retraction velocity $\Cas$ with the control parameter $\Ohs$ in the viscous limit by following the location and magnitude of the local rate of viscous dissipation during Taylor-Culick retractions in viscous surroundings. 

\section{Conclusion and outlook} \label{sec:conclusion}

In this paper, we have studied the effects of the surrounding media on the retraction dynamics of liquid sheets in three canonical configurations. In the \emph{classical} Taylor-Culick configuration, the interplay between capillarity and inertia of the film results in a constant retraction velocity. We can further neglect the surrounding medium as it does not influence the retraction process. However, for a film retracting in a dense and viscous oil (\emph{two-phase} configuration), and that at an oil-air interface (\emph{three-phase}), both inertia and viscosity of the oil phase influence the retraction process. The former presents itself as an added mass-like effect. Even though capillarity still governs the constant retraction velocity, the surrounding medium's inertia reduces the magnitude of the film's momentum as it retracts.

Moreover, when the viscosity of the oil is significantly higher than that of the film, the viscous stresses dictate the retraction process and set the velocity scale. To further demystify the energy balance in this process, we used thermodynamically consistent energy transfer mechanisms to understand the fate of the released surface energy owing to the loss of surface area of the retracting film. This energy is injected into the system and manifests itself as kinetic energy and viscous dissipation. In the inertial regime, the proportions of kinetic energy and viscous dissipation are the same, conforming to the analyses of \citet{culick-1960-japplphys}. However, in the viscous regime, the total surface energy released goes into viscous dissipation in the surroundings. 

Following the lumped elements analysis, motivated by \citet{taylor-1959-procrsoclonda, culick-1960-japplphys}, we also developed scaling relations to relate the non-dimensionalized retraction velocity ($\Wef$ and $\Cas$) with the control parameter $\Ohs$. In the inertial limit, the Weber number $\Wef$ based on the retraction velocity is a constant for all three configurations. On the other hand, in the viscous limit, the retraction velocity in the \emph{two-phase} configuration scales with the visco-capillary velocity scale (constant capillary number, $\Cas \sim \mathcal{O}\left(1\right)$); while for the \emph{three-phase} configuration, the capillary number $\Cas$ increases with increasing $\Ohs$, owing to the localization of viscous dissipation near the three-phase contact line. 

A natural extension of the present work would be to understand the retraction of non-Newtonian sheets and filaments \citep{sen_lohse_2021} in similar surroundings. In such scenarios, the retraction dynamics will depend not only on capillarity and viscosity as described in this work, but also on the rheological properties of both the film and the surroundings. Furthermore, in a broader perspective, the precursor film-based three-fluid volume of fluid method can be used to elucidate several spreading phenomena, both at small and large scales, e.g., drop-film interactions in the inkjet printing process \citep{lohse2022fundamental} and late time spreading during oil spillage \citep{hoult1972oil}, respectively.\\

\noindent{\textbf{Acknowledgments}}

We acknowledge Pim Dekker for carrying out initial experiments. We would like to thank Maziyar Jalaal, Jacco Snoeijer, and Andrea Prosperetti for discussions. This work was carried out on the national e-infrastructure of SURFsara, a subsidiary of SURF cooperation, the collaborative ICT organization for Dutch education and research. \\
 
 \noindent{\textbf{Funding}}
 
We acknowledge the funding by the ERC Advanced Grant No. 740479-DDD, an Industrial Partnership Programme of the Netherlands Organisation for Scientific Research (NWO), cofinanced by Canon Production Printing B. V., University of Twente, and Eindhoven University of Technology, and the Max Planck Center Twente. \\

\noindent{\textbf{Declaration of interests}}

The authors report no conflict of interest. \\

\noindent{\textbf{Supplementary information}} \label{SM}

Supplementary information is available at (URL to be inserted by publisher). \\

\noindent{\textbf{Author ORCID}}

V. Sanjay \href{https://orcid.org/0000-0002-4293-6099}{https://orcid.org/0000-0002-4293-6099};  

U. Sen \href{https://orcid.org/0000-0001-6355-7605}{https://orcid.org/0000-0001-6355-7605};  

P. Kant \href{https://orcid.org/0000-0001-8177-0848}{https://orcid.org/0000-0001-8177-0848};  

D. Lohse \href{https://orcid.org/0000-0003-4138-2255}{https://orcid.org/0000-0003-4138-2255}. \\

\appendix

\section{Classical Taylor-Culick retractions}\label{App::ClassicalTC}

In this section, we discuss the classical Taylor-Culick retractions, which is modeled using the numerical method used for the two-phase configuration (see \S~\ref{sec:2-phaseTC Num setup}) by replacing the surrounding medium ($s$) with air ($a$). The volume of fluid (VoF) tracer advection equation (\ref{Eqn::Vof1}), and the \citet{brackbill1992continuum} surface tension force formulation (\ref{Eqn::SurfaceTension1}) remain the same, whereas, the VoF property equations are modified as

\begin{align}
	\label{Eqn::density1}
	\tilde{\rho} &= \Psi + \left(1-\Psi\right)\frac{\rho_{a}}{\rho_{f}},\\
	\label{Eqn::Oh1}
	Oh &= \Psi \Ohf + \left(1-\Psi\right)\Oha,
\end{align}

\noindent where $\rho_a/\rho_f$ is the air to film density ratio (fixed at $10^{-3}$), and the two dimensionless groups

\begin{align}
	\label{Eqn::Ohfa}
	\Ohf = \frac{\eta_f}{\sqrt{\rho_f\left(2\gammaaf\right)h_0}}, \quad \Oha = \frac{\eta_a}{\sqrt{\rho_f\left(2\gammaaf\right)h_0}}
\end{align}

\noindent  represent the film Ohnesorge number and the air Ohnesorge number (fixed at $10^{-5}$), respectively. 

\begin{figure}
	\centering
	\includegraphics[width=\textwidth]{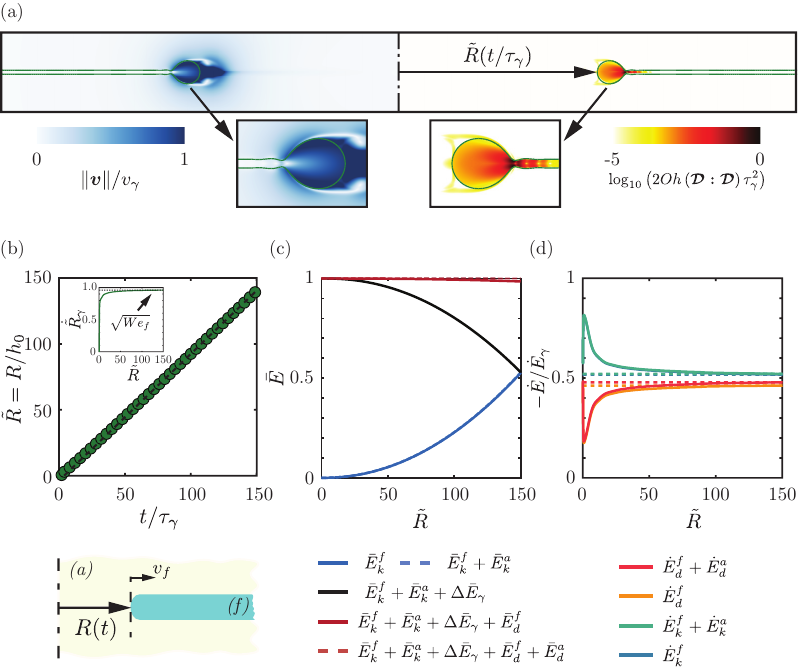}	
	\caption{Classical Taylor-Culick retractions: (a) The morphology of the flow when the dimensionless hole radius $\tilde{R} = 50$. The left hand side contour shows the velocity magnitude normalized with the inertio-capillary velocity scale ($\|\boldsymbol{v}\|/v_\gamma$), while the right hand side shows the dimensionless rate of viscous dissipation per unit volume normalized using the inertio-capillary scales $\left(2Oh\left(\boldsymbol{\mathcal{D}:\mathcal{D}}\right)\tau_\gamma^2\right)$, represented on a $\log_{\text{10}}$ scale to differentiate the regions of maximum dissipation. (b) Temporal evolution of $\tilde{R}(t)$. Time is normalized using the inertio-capillary time scale, $\tau_\gamma = \sqrt{\rho_f h_0^3/\gammasf}$. Inset of panel (b) shows the variation of dimensionless growth rate of the hole radius. Notice that $\sqrt{\Wef} = \lim\limits_{\tilde{R} \to \infty}\dot{\tilde{R}}_\gamma = 1$. (c) Energy budget where the energies ($E$) are normalized using the total surface energy released as the film retracts, creating a hole of radius $\tilde{R}_{\text{max}} = 150$. (d) Variations of the rate of change of energy $\dot{E}(t)$ as a fraction of the rate of energy injection into the system ($-\dot{E}_\gamma (t)$) with dimensionless hole radius $\tilde{R}(t)$. The superscripts account for the film ($f$) and air ($a$). The Ohnesorge number of the film for this simulation is $\Ohf = 0.05$, and that of air is $\Oha = 10^{-5}$ to respect the assumption that the surrounding medium has negligible effect on the retraction process \citep{taylor-1959-procrsoclonda, culick-1960-japplphys}. Additionally, the air-to-film density ratio is $\rho_a/\rho_f = 10^{-3}$. Also see supplementary movie SM4.}
	\label{fig:ClassicalTC}
\end{figure}

Figure~\ref{fig:ClassicalTC} summarizes the results of the classical Taylor-Culick retractions for a typical $\Ohf = 0.05$. After the initial transients, the growing hole follows a linear evolution in time and the growth rate approaches the Taylor-Culick velocity (\ref{eq:v_TC}), see figure~\ref{fig:ClassicalTC}b and its inset). In the steady state, both the water film and the ambient air move (figure~\ref{fig:ClassicalTC}a), but the density of air is negligible as compared to that of the film. Consequently, the air does not contribute to the force or energy equilibrium described below. 

\subsection{Force balance}
For the classical configuration (figure~\ref{fig:configs}a), the force balance strictly implies that the capillary force ($F_{\gamma}(t)$) equals the rate of change of momentum ($P(t)$) of the moving rim written as \citep{taylor-1959-procrsoclonda}

\begin{align}
	F_{\gamma}(t) = \frac{dP(t)}{dt} = \frac{d}{dt}\left(m(t)v_f\right),
	\label{eq:classical balance 1}
\end{align}

\noindent where the capillary force is given by $F_\gamma(t) = 2\gammaaf\left(2\pi R(t)\right)$, $\gammaaf$ is the surface tension coefficient between the film and the surrounding air. Assuming that the film velocity $v_f$ is a constant, we can simplify (\ref{eq:classical balance 1}) to

\begin{align}
	2\gammaaf\left(2\pi R(t)\right) = v_f\frac{dm(t)}{dt},
	\label{eq:classical balance 2}
\end{align}

\noindent where we can employ the continuity equation to get
\begin{align}
	\frac{dm(t)}{dt} = \rho_f v_f h_0\left(2\pi R(t)\right).
	\label{eq:classical balance 2b}
\end{align}

\noindent Further, using (\ref{eq:classical balance 2}) and (\ref{eq:classical balance 2b}),

\begin{align}
	2\gammaaf\left(2\pi R(t)\right) = \rho_fv_f^2h_0\left(2\pi R(t)\right),
	\label{eq:classical balance 3}
\end{align}

\noindent for the classical configuration (figure~\ref{fig:configs}a), giving

\begin{align}
	v_f = \sqrt{ \frac{2\gammaaf}{\rho_f h_0}}.
	\label{eq:classical v}
\end{align}

\noindent Note that (\ref{eq:classical balance 1})--(\ref{eq:classical balance 3}) are similar to the calculations of \citet{taylor-1959-procrsoclonda}, and only considers momentum equilibrium while disregarding the fate of the liquid accumulated in the rim \citep{villermaux2020fragmentation}. Furthermore, it assumes no interaction with the surrounding medium (air). In terms of the dimensionless numbers, (\ref{eq:classical v}) implies that $\Wef~=~\rho_fv_f^2h_0/(2\gammaaf)$ is constant and equal to 1 (i.e., $v_f = v_{\text{TC}}$, see (\ref{eq:v_TC})).

\subsection{Energy balance}
\citet{dupre1867theorie, dupre1869theorie} wrongly assumed that the entire surface energy released during the retraction manifests as the kinetic energy of the film \citep{rayleigh-1891-nature}, giving

\begin{align}
	-\dot{E}_\gamma(t) &= \dot{E}_k^f(t),\\
	2\gammaaf\left(2\pi R(t)\right)v_f &= \frac{d}{dt}\left(\frac{1}{2}m(t)v_f^2\right).
\end{align}

\noindent Using conservation of mass $dm(t) = \rho vh_0\left(2\pi R(t)\right)dt$, \citet{dupre1867theorie, dupre1869theorie} calculated the retraction velocity to be

\begin{align}
	v_f = \sqrt{\frac{4\gammaaf}{\rho_f h_{0}}} = \sqrt{2}v_{\text{TC}},
\end{align}

\noindent which is off by a factor of $\sqrt{2}$ \citep[see Dupr{\'e}-Rayleigh paradox in][]{villermaux2020fragmentation}. 

\citet{culick-1960-japplphys} realized that the correct energy balance entails that the rate of surface energy released should be distributed equally into an increase in kinetic energy of the rim and the viscous dissipation inside the film (\ref{Eqn::CulickEnergyBalance1}). Figures~\ref{fig:ClassicalTC}c and d illustrate the energy balance associated with the classical Taylor-Culick retractions (note that $\bar{E}_k^f(t) \approx \bar{E}_d^f(t)$ in figure~\ref{fig:ClassicalTC}c and $\dot{E}_k^f(t) \approx \dot{E}_d^f(t)$ in figure~\ref{fig:ClassicalTC}d). 

\begin{align}
	\label{Eqn::CulickEnergyBalance1}
	-\dot{E}_\gamma(t) = \dot{E}_k^f(t) + \dot{E}_d^f(t),
\end{align}

\noindent where $-\dot{E}_\gamma (t) \approx 2\gammaaf\left(2\pi R(t)\right)v_f$ \citep[see appendix~\ref{App::EnergyBalance} and][]{bohr2021surface}. Note that the rate of viscous dissipation at any given instant is analogous to the inelastic collision of a tiny fluid parcel in the film with the massive rim. Indeed, the local viscous dissipation $\left(2Oh\left(\boldsymbol{\mathcal{D}:\mathcal{D}}\right)\tau_\gamma^2\right)$ is maximum in the region connecting the rim to the film (figure~\ref{fig:ClassicalTC}a). Consequently \citep{culick-1960-japplphys},  

\begin{align}
	2\gammaaf\left(2\pi R(t)\right)v_f = \frac{d}{dt}\left(\frac{1}{2}m(t)v_f^2\right) + \frac{1}{2}\frac{dm(t)}{dt}v_f^2.
	\label{Eqn::CulickEnergyBalance2}
\end{align}

\noindent Again, using conservation of mass $dm = \rho vh_0\left(2\pi R(t)\right)dt$ and rearranging (\ref{Eqn::CulickEnergyBalance2}), we get
\begin{align}
	v_f = v_{\text{TC}} = \sqrt{\frac{2\gammaaf}{\rho_f h_{0}}}
	\label{Eqn::TC-velocity-classical App}
\end{align}

\noindent for the classical configuration. 

\section{Energy calculations}\label{App::EnergyBalance}
This appendix explains the motivation and mathematical expressions used in the present study to describe different energy transfers, and their rates, as discussed in \S~\ref{sec:energetics}. Similar approaches have been used in the literature to study the dynamics of two-phase flows \citep{sanjay_lohse_jalaal_2021, bohr2021surface}. Here, we extend these formulations to three-phase flows.  

The kinetic energies and viscous dissipations associated with the three fluids are given by \citep[p.~50-51]{landau2013course}

\begin{align}
	\label{Eqn::Ek}
	E_k^j &= \frac{1}{2}\rho_j\int_{\Omega_j}\|\boldsymbol{u}\|^2\,\mathrm{d}\Omega_j,\\
	\label{Eqn::Ed}
	E_d^j &= 2\int_t\left(\int_{\Omega_j}\eta_j\left(\boldsymbol{\mathcal{D}}:\boldsymbol{\mathcal{D}}\right)\,\mathrm{d}\Omega_j\right)\mathrm{d}t = \int_t\left(\int_{\Omega_j}\varepsilon_j\,\mathrm{d}\Omega_j\right)\mathrm{d}t.
\end{align}

\noindent where $\mathrm{d}\Omega_j$ is the differential volume element associated with the $j^{\text{th}}$ fluid. Additionally, $\rho_j$ and $\eta_j$ denote the density and viscosity, respectively, of the $j^{\text{th}}$ fluid. In the present work,  $j = f$ (film, water), $s$ (surroundings, oil), and $a$ (air). Furthermore, in terms of rates, 

\begin{align}
	\label{Eqn::Ekt}
	\frac{dE_k^j}{dt} &= \frac{d}{dt}\left(\frac{1}{2}\rho_j\int_{\Omega_j}\|\boldsymbol{u}\|^2\,\mathrm{d}\Omega_j\right),\\
	\label{Eqn::Edt}
	\frac{dE_d^j}{dt} &= \int_{\Omega_j}\varepsilon_j\,\mathrm{d}\Omega_j.
\end{align}

Next, the total surface energy $E_\gamma$ of the system for the three-phase configuration is

\begin{align}
	\label{Eqn::Es1}
	E_\gamma = \int_{\mathcal{A}_{sf}}\gammasf\mathrm{d}\mathcal{A}_{sf} + \int_{\mathcal{A}_{sa}}\gammasa\mathrm{d}\mathcal{A}_{sa},
\end{align}

\noindent where $\gamma_{ij}$ and $\mathcal{A}_{ij}$ are the interfacial tension coefficient and area, respectively, associated with an interface between the $i^{\text{th}}$ and the $j^{\text{th}}$ fluids. Note that, the assumption of a precursor film of oil (surroundings, $s$) on the water film ($f$) implies that there is no film-air interface. Additionally, $\gamma_{sf} = 2\gamma_{sa}$ (see \S~\ref{sec:3-phaseTC Num methods}). 

\begin{align}
	\label{Eqn::Es2}
	E_\gamma = \gammasf\left(\mathcal{A}_{sf} + \mathcal{A}_{sa}/2\right),
\end{align}

\noindent So, the rate of surface energy released during the retraction process in the three-phase configuration is

\begin{align}
	\label{Eqn::Est1}
	\dot{E}_\gamma = \gammasf\left(\dot{\mathcal{A}}_{sf} + \dot{\mathcal{A}}_{sa}/2\right),
\end{align}

\noindent where $\dot{\mathcal{A}}_{ij}$ is the rate of change of interfacial area. 

For the two-phase configuration, there is no air ($\mathcal{A}_{sa} = 0$), and the rate of change of surface energy is simply

\begin{align}
	\label{Eqn::Est2}
	\dot{E}_\gamma = \gammasf\dot{\mathcal{A}}_{sf}.
\end{align}

\begin{figure}
	\centering
	\includegraphics[width=\textwidth]{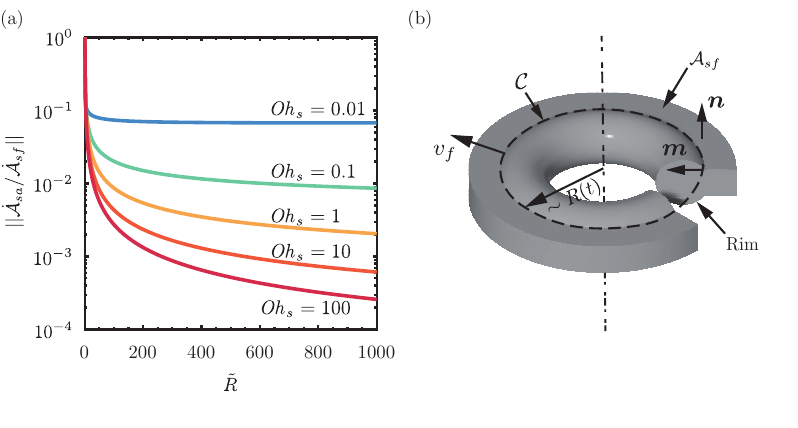}	
	\caption{(a) Variation of the ratio of the magnitudes of the rate of change of surroundings-air interfacial area ($\dot{\mathcal{A}}_{sa}$) to that of the surroundings-film ($\dot{\mathcal{A}}_{sf}$) with the dimensionless hole radius $\tilde{R}(t) = R(t)/h_0$. (b) Schematic showing the control surface $\mathcal{A}_{sf}$ (free surface of the film without the rim) used for the calculation of the rate of change of surface energy.}
	\label{fig:AreaRateRatio}
\end{figure}

Note that we use (\ref{Eqn::Ed} -- \ref{Eqn::Est2}) for calculating the energies, and their rates of change, in figures~\ref{fig:energy},~\ref{fig:energy_rate}, and~\ref{fig:ClassicalTC}. However, to better understand the individual contributions of the two terms on the right hand side of (\ref{Eqn::Est1}), figure~\ref{fig:AreaRateRatio} illustrates the ratio of the rate of change of surroundings-air interfacial area ($\dot{\mathcal{A}}_{sa}$) to that of the surroundings-film ($\dot{\mathcal{A}}_{sf}$). Initially, at very small hole radii ($\tilde{R} \to 0$), the two rates are comparable ($\dot{\mathcal{A}}_{sa} \sim \dot{\mathcal{A}}_{sf}$). But, after these initial transients, the rate of change in the surroundings-film interface area dominates ($\dot{\mathcal{A}}_{sf} \gg \dot{\mathcal{A}}_{sa}$).  Therefore, even for the three-phase configuration, in the steady state,

\begin{align}
	\label{Eqn::Est3}
	\dot{E}_\gamma \approx \gammasf\dot{\mathcal{A}}_{sf}.
\end{align}

As a result, we only need to evaluate $\dot{\mathcal{A}}_{sf}$ for developing a scaling for the rate of change of surface energy. For doing this, we use the analysis presented in \cite{bohr2021surface}, written in our notations as 

\begin{align}
	\label{Eqn::Gammadot1}
	\dot{\mathcal{A}}_{sf} = \int_{\mathcal{A}_{sf}} \kappa\left(\boldsymbol{U}\cdot\boldsymbol{n}\right)\mathrm{d}\mathcal{A}_{sf} + \int_{\mathcal{C}}\left(\boldsymbol{U}\cdot\boldsymbol{m}\right)\mathrm{d}\mathcal{C}
\end{align} 

\noindent for a control volume bounded by the control surface $\mathcal{A}_{sf}$ (free surface of the film without the rim, figure~\ref{fig:AreaRateRatio}b). Here, $\boldsymbol{U}$ is the velocity of differential control volume bounded by $\mathrm{d}\mathcal{A}_{sf}$, $\kappa$ the curvature at this location, and $\boldsymbol{n}$ is a unit vector normal to $\mathrm{d}\mathcal{A}_{sf}$. Lastly, the control surface $\mathcal{A}_{sf}$ is bounded by the contour $\mathcal{C}$, and $\boldsymbol{m}$ is a unit vector perpendicular to this contour. Note that capillary traction acts perpendicular to $\mathcal{C}$ away from the axis of symmetry. The first term on the right hand side of (\ref{Eqn::Gammadot1}) accounts for the change in surface area due to inflation normal to $\mathcal{A}_{sf}$, and the second term is a consequence of the distortion of $\mathcal{A}_{sf}$ in the tangential direction, i.e., stretching, or in this case, compression (the growing hole). With this choice of the control surface, the dilation normal to $\mathcal{A}_{sf}$ is zero (area inflation only occurs at the rim which we ignore), and

\begin{align}
	\label{Eqn::Gammadot2}
	\dot{\mathcal{A}}_{sf}(t) \approx \int_{\mathcal{C}(t)}\left(\boldsymbol{U}\cdot\boldsymbol{m}\right)\mathrm{d}\mathcal{C} = -2v_f\left(2\pi R(t)\right),
\end{align} 

\noindent where the factor $2$ comes in because of the two surfaces (top and bottom). Therefore, for both the two-phase as well as three-phase Taylor-Culick retractions, the rate of injection of energy in the system is

\begin{align}
	\label{Eqn::Egammadot3}
	-\dot{E}_\gamma(t) \approx 2\gammasf v_f\left(2\pi R(t)\right).
\end{align} 

\noindent Note that while calculating the rate of change of surface energy, we did not account for the growth of the rim because it is much slower than the growth of the hole, and the flow is predominantly in the radial direction \citep[see figures~\ref{fig:two-phaseTemporal} and~\ref{fig:three-phaseTemporal}, and][]{gordillo2011asymptotic}.

\section{Code availability}
The codes used in the present article are permanently available at \citet{basiliskVatsal}.

\bibliographystyle{jfm}
\bibliography{ThreePhaseTaylorCulick}

\end{document}